\documentclass[twocolumn,prd,superscriptaddress,showpacs,floatfix,%
preprintnumbers, nofootinbib,hyperref]{revtex4}  
\usepackage{epsfig}   
\usepackage{bm}  
\usepackage{color}

\newcommand{\be}{\begin{eqnarray}}  
\newcommand{\ee}{\end{eqnarray}}

\begin{document}  

\title{Multi-angle effects in self-induced  oscillations for different supernova neutrino fluxes}

\author{Alessandro Mirizzi} 
\affiliation{II Institut f\"ur Theoretische Physik, Universit\"at Hamburg, Luruper Chaussee 149, 22761 Hamburg, Germany}

\author{Ricard Tom{\`a}s}   
\affiliation{II Institut f\"ur Theoretische Physik, Universit\"at Hamburg, Luruper Chaussee 149, 22761 Hamburg, Germany}


\begin{abstract}   
 The non-isotropic nature of the neutrino emission from a supernova (SN) core
 might potentially affect the flavor evolution 
of the  neutrino ensemble, via neutrino-neutrino interactions in the deepest SN regions. We investigate the dependence of these 
``multi-angle effects" on  the original SN neutrino fluxes in a three-flavor framework.
We  show  that the pattern of the spectral crossings
 (energies where $F_{\nu_e}= F_{\nu_x}$, and  $F_{\overline\nu_e}= F_{\overline\nu_x}$)
 is crucial in determining the impact of multi-angle effects on the flavor evolution.
For neutrino spectra presenting only a single-crossing, 
synchronization of different angular modes  prevails over 
multi-angle effects, producing  the known ``quasi single-angle'' evolution. 
Conversely,  in the presence of spectra with  multiple crossing energies,
synchronization is not stable.
 In this situation,  
multi-angle effects would produce a  sizable delay in  the onset of the flavor  conversions, as recently observed. We show that, due to the only partial 
  adiabaticity of the  evolution at large radii, the multi-angle 
suppression can be so strong  to dramatically affect
the final oscillated neutrino spectra.
 In particular three-flavor effects, associated with the solar parameters, could be 
washed-out in multi-angle simulations.
\end{abstract}   
   
\pacs{14.60.Pq, 97.60.Bw}   
   
\maketitle

\section{Introduction}         \label{intro}  
 
 The characterization of the flavor conversions for neutrinos emitted by a 
 stellar   collapse is a field of intense activity. In particular,
the flavor transformation probabilities in supernovae (SNe) not only depend on the matter  
background~\cite{Matt,Dighe:1999bi}, but also on the neutrino fluxes themselves:  
neutrino-neutrino interactions provide a nonlinear term in the  
equations of motion~\cite{Pantaleone:1992eq, Sigl:1992fn,McKellar:1992ja} that causes  
collective flavor transformations~\cite{Qian:1994wh,Samuel:1993uw,  
  Kostelecky:1993dm, Kostelecky:1995dt, Samuel:1996ri, Pastor:2001iu,  
  Wong:2002fa, Abazajian:2002qx, Pastor:2002we, Sawyer:2004ai, Sawyer:2005jk}.
 Only  
recently~\cite{Duan:2005cp,Duan:2006an,Hannestad:2006nj} it has  been fully appreciated that in the SN context these  
collective effects give rise to qualitatively new  
phenomena
(see, e.g.,~\cite{Duan:2010bg} for a recent review). The main consequence of 
this unusual type of  flavor transitions is 
an exchange of the spectrum of the electron species $\nu_e$ (${\bar\nu}_e$) with
the non-electron ones  $\nu_x$ (${\bar\nu}_x$) in certain energy intervals. 
These flavor exchanges are called ``swaps'' marked by the  ``splits'', which are
the boundary features at the edges of each swap interval~\cite{Duan:2006an,Duan:2010bg, Fogli:2007bk,Fogli:2008pt, Raffelt:2007cb, 
Raffelt:2007xt, Duan:2007bt, Duan:2008za, Gava:2008rp, Gava:2009pj,Dasgupta:2009mg,Friedland:2010sc,Dasgupta:2010cd}. The location and the  number of these splits, as well as their dependence on the neutrino mass hierarchy, is crucially dependent on the flux ordering among different neutrino species~\cite{Fogli:2009rd,Choubey:2010up}. 

In this context, one of the main complication in the simulation of the flavor evolution
is that the flux of neutrinos emitted from a 
 supernova core is far from isotropic. The current-current nature of the weak-interaction Hamiltonian implies that the interaction energy between neutrinos of 
momenta ${\bf p}$ and ${\bf q}$ is proportional to 
$(1-{\bf v}_{\bf p} \cdot {\bf v}_{\bf q})$, where ${\bf v}_{\bf p}$
is the neutrino velocity~\cite{Qian:1994wh,Pantaleone:1992xh}. In a non-isotropic medium this velocity-dependent term would not average to zero, producing a different 
refractive index for neutrinos propagating on different trajectories. 
This is the origin of the so-called ``multi-angle effects''~\cite{Duan:2006an}, which hinder the maintenance    of the coherent oscillation 
behavior for different neutrino modes~\cite{Raffelt:2007yz,Fogli:2007bk,EstebanPretel:2007ec,Sawyer:2008zs}. In~\cite{Raffelt:2007yz} it has been shown that in
a dense neutrino gas initially composed of only $\nu_e$ and $\overline\nu_e$ with 
equal fluxes, multi-angles effects would  rapidly lead to flavor decoherence, resulting 
in flux equilibration among electron and non-electron (anti)neutrino species.
On the other hand, in the presence of relevant flavor asymmetries
between $\nu_e$ and ${\overline \nu}_e$
 multi-angle effects can be 
suppressed. 
In particular, during the early SN accretion phase, 
one expects as neutrino flux ordering $\Phi_{\nu_e}\gg  \Phi_{\overline\nu_e}\gg  \Phi_{\nu_x}= \Phi_{\overline\nu_x}$
~\cite{Raffelt:2003en, Fischer:2009af, Huedepohl:2009wh}, defined in terms of the total 
neutrino number fluxes  $\Phi_\nu$ for the different flavors.
This case  would practically correspond to neutrino spectra
with a  \emph{single crossing} point at $E \to \infty$  (where $F_{\nu_e}= F_{\nu_x}$, and  $F_{\overline\nu_e}= F_{\overline\nu_x}$)
 since   $F_{\nu_e}(E) >  F_{\nu_x}(E)$ for all the relevant
energies (and analogously for $\bar\nu$).  
For spectra with a single crossing, it has been shown in~\cite{EstebanPretel:2007ec} that  the presence of
significant flavor asymmetries between $\nu_e$ and $\overline\nu_e$ fluxes
 guarantees the synchronization of different angular modes at low-radii ($r\lesssim 100$~km), so that essentially nothing happens close to the neutrinosphere because the in-medium mixing is extremely
small. Therefore, the possible onset of multi-angle effects is  delayed after the synchronization phase. 
Then,  the flavor evolution is adiabatic to produce spectral splits and swaps but not enough to allow the multi-angle instability to grow and  produce significant decoherence effects. Therefore, the resultant neutrino flavor conversions would be described by an effective ``quasi single-angle'' behavior. In this case, the self-induced spectral swaps and splits would be only marginal smeared by
multi-angle effects, as explicitly shown in~\cite{Duan:2006an,Fogli:2007bk,Fogli:2008pt}.
This reassuring result has been taken as granted in most of the  further studies that typically adopted the averaged single-angle
approximation.

However, this nice picture does not represent the end of the story for multi-angle effects.%
\footnote{
We mention that recently multi-angle effects have been included also in the study of
the flavor evolution of the $\nu_e$ neutronization burst in O-Ne-Mg supernovae.
We address the interested reader to~\cite{Cherry:2010yc}.}
A different result has been recently shown in~\cite{Duan:2010bf}, where 
multi-angle effects are explored, assuming a flux ordering of the 
type $\Phi_{\nu_x} \gtrsim \Phi_{\nu_e} \gtrsim \Phi_{\overline\nu_e}$, possible during the SN cooling phase, where
one expects a moderate flavor hierarchy among different species and a ``cross-over'' among non-electron and electron species
is possible~\cite{Raffelt:2003en, Fischer:2009af, Huedepohl:2009wh}. 
This case  would correspond to neutrino spectra
with \emph{multiple crossing} points, i.e.  with $F_{\nu_e}(E) >  F_{\nu_x}(E)$ at lower energies, and $F_{\nu_e}(E) <  F_{\nu_x}(E)$
 at higher energies (and analogously for $\bar\nu$). 
For such spectra, it has been shown in~\cite{Raffelt:2008hr} that the synchronization is   not a stable solution
for a neutrino gas in presence of a large neutrino density.  Therefore, collective flavor
conversions would be possible at low-radii in the single-angle scheme~\cite{Raffelt:2008hr}, in a region where
one would have naively expected synchronization.
 However,
it has been shown in~\cite{Duan:2010bf} that the presence of a large dispersion in the neutrino-neutrino refractive index, induced by multi-angle effects, seems to block the development of these collective flavor conversions close to the neutrinosphere.
This recent result extends  the  finding  obtained with a toy model in~\cite{Raffelt:2008hr}. The delay of the self-induced
flavor conversions for this case is also visible in the multi-angle
simulations  in~\cite{animations}.
So, it is apparent that multi-angle effects are relevant not only for fluxes with 
small flavor asymmetries, where they trigger a quick flavor \emph{decoherence}, but also in cases of spectra with multiple crossing points, where
  multi-angle effects can \emph{suppress} flavor conversions at low-radii.  
 
 Triggered by the contrasting impact of the multi-angle 
 effects for fluxes typical of the accretion and cooling phase, we take a closer 
 look at the dependence of these effects on the neutrino flux ordering.  
The plan of our work is as follows.
 In Sec.~II we  introduce our supernova flux models, and describe the equations for the flavor
conversions  in the multi-angle and single-angle case. In Sec.~III we show and explain our numerical results for the single-angle and
multi-angle flavor evolution for some representative choices of SN neutrino fluxes. 
In particular, we select three cases corresponding respectively to $a)$ 
single-crossed neutrino spectrum with $\Phi_{\nu_e}\gg  \Phi_{\overline\nu_e}\gg  \Phi_{\nu_x}$, producing  a ``quasi single-angle'' flavor 
evolution, $b)$ multiple-crossed spectrum with 
$\Phi_{\nu_x} \gtrsim \Phi_{\nu_e} \gtrsim \Phi_{\overline\nu_e}$,
where single-angle and multi-angle evolutions give
significantly different final neutrino spectra, $c)$ small flavor asymmetries, i.e.
$\Phi_{\nu_e} \approx \Phi_{\overline\nu_e}\approx\Phi_{\nu_x}$, where the  
 multi-angle suppression is small, and multi-angle decoherence produces a  partial flavor equilibration among the different species.
Finally,
in Sec.~IV we draw inferences from our results and summarize.
Technical aspects are discussed in the Appendix.


\section{Supernova neutrino  fluxes and   equations of motion} 

\subsection{Supernova flux models} 

In the presence of neutrino-neutrino interactions the flavor conversions for SN neutrinos are described by non-linear
equations. Therefore, SN neutrino oscillations will crucially depend on the initial neutrino fluxes.
We define  $F_{\nu_\alpha}$ as the   number flux of a given neutrino species $\nu_{\alpha}$ 
 emitted with energy $E$ in any direction at the neutrinosphere.
 In a
supernova $\nu_e$ and ${\bar\nu}_e$ are distinguished from other flavors
due to their  charged-current interactions. The
$\nu_\mu$, $\nu_\tau$ and their antiparticles, on the other hand, are
produced at practically identical rates. Following the
standard terminology~\cite{Dasgupta:2007ws}, we define the two relevant non-electron
flavor states as $\nu_{x,y} = \cos \theta_{23}\nu_\mu \mp \cos \theta_{23}\nu_\tau$, where
$\theta_{23}\simeq\pi/4$ is the atmospheric mixing angle. Since the
initial $\nu_x$ and $\nu_y$ fluxes are identical, the primary neutrino
fluxes are best expressed in terms of $\nu_e$, ${\bar\nu}_e$ and $\nu_x$.
 For half-isotropic emission these three relevant SN $\nu$ original   neutrino number fluxes
for the different species 
are given by~\cite{Fogli:2007bk}
\begin{equation}  
\label{Ybeta} F_{\nu_\alpha}^0(E)= {\Phi^0_{{\nu_\alpha}}}\,\varphi_{\nu_\alpha}(E)\ ,  
\end{equation}  
where  
\begin{equation}
\Phi_{{\nu_\alpha}}^0 = \frac{1}{4 \pi^2 R^2}\frac{L_{{\nu_\alpha}}}{\langle E_{\nu_\alpha}\rangle}
\end{equation}
 is the total  number flux at the neutrinosphere radius $R$, that in our numerical examples we will take
$R=10$~km. The neutrino luminosity is $L_{{\nu_\alpha}}$ and the neutrino average energy $\langle E_{\nu_\alpha}\rangle$.
The function  $\varphi_{\nu_\alpha}(E)$ is the normalized neutrino spectrum ($\int dE  
\; \varphi_{\nu_\alpha}=1$) and parametrized as~\cite{Keil:2002in}: 
\begin{equation}  
\label{varphi} \varphi_{\nu_\alpha}(E)=\frac{\beta^\beta}{  
\Gamma(\beta)}\frac{E^{\beta-1}}{\langle E_{\nu_\alpha} 
\rangle^{\beta}} e^{-\beta E/\langle  
E_{\nu_\alpha}\rangle}\; ,  
\end{equation}  
where $\beta$  is a spectral parameter, and
 $\Gamma(\beta)$ is the Euler gamma function. 
The values of the parameters are model dependent (e.g. see the Fig.~3 in~\cite{Fogli:2009rd}). For our numerical illustrations, we choose  
\begin{equation}  
\label{Eave} (\langle E_{\nu_e} \rangle,\;\langle E_{\bar\nu_e}  
\rangle,\;\langle E_{\bar\nu_x}  
\rangle )  
 = (12,\;15,\;18)\ \mathrm{MeV}\ ,  
\end{equation}  
and $\beta=4$, from the admissible parameter ranges~\cite{Keil:2002in}. 

We will consider three representative cases for the  ratios of the neutrino fluxes, namely 
\begin{eqnarray}
  \Phi^0_{\nu_e}\;:\;\Phi^0_{{\bar{\nu}}_e}\;:\;\Phi^0_{\nu_x} &=& 2.40\;:\;1.60\;:\;1.0 \,\ , \nonumber \\
  \Phi^0_{\nu_e}\;:\;\Phi^0_{{\bar{\nu}}_e}\;:\;\Phi^0_{\nu_x} &=& 0.85\;:\;0.75\;:\;1.0 \,\ , \nonumber \\
  \Phi^0_{\nu_e}\;:\;\Phi^0_{{\bar{\nu}}_e}\;:\;\Phi^0_{\nu_x} &=& 0.81\;:\;0.79\;:\;1.0 \,\ .
  \label{eq:cases}
\end{eqnarray}
The first case  represents a flux ordering with a  $\nu_e$ dominance, typical of the accretion phase, practically producing  a single-crossed spectrum.
The other two cases represent fluxes possible during the cooling phase, with a moderate $\nu_x$ dominance and different flavor asymmetries,
producing a  multiple-crossed spectrum.
We will see how multi-angle effects would have a different impact for these three
cases.

\subsection{Equations of motion}

Mixed neutrinos are described by matrices of density
$\rho_{\bf p}$ and ${\bar \rho}_{\bf p}$ for each (anti)neutrino mode. The diagonal
entries are the usual occupation numbers whereas the off-diagonal
terms encode phase information. 
We are studying the spatial evolution of the neutrino
fluxes in a quasi-stationary situation. Therefore, the matrices $\rho_{\bf p}$
do not explicitly depend on time, so that the 
evolution reduces to the Liouville
term involving only spatial derivatives.  Moreover, we assume spherical
symmetry so that the only relevant spatial variable
is the radial coordinate $r$. In this case, the explicit form of the equations
of motion (EoMs)
has been obtained in~\cite{Sigl:1992fn,McKellar:1992ja,Hannestad:2006nj,EstebanPretel:2007ec,Dasgupta:2008cu,Duan:2008fd}. 

\begin{equation}
\textrm{i} {\bf v}_{\bf p} \cdot {\nabla}_r \rho_{\bf p} = [{\sf H}_{\bf p}, \rho_{\bf p}] \,\ ,
\label{eomMA}
\end{equation}
where ${\bf v}_{\bf p}$ is the velocity and the Hamiltonian reads 
\begin{equation}
{\sf H}_{\bf p} = {\sf\Omega}_{\bf p} + {\sf V}_{\rm MSW} +
{\sf V}_{\nu\nu} \,\ .
\label{eq:Ham}
\end{equation}
In a three flavor scenario, 
the matrix of the vacuum oscillation frequencies for neutrinos is ${\sf \Omega}_{\bf p}= \textrm{diag}(m_1^2,m_2^2,m_3^2)/2|{\bf p}|$ in the mass basis.
For antineutrinos ${\sf \Omega}_{\bf p} \to -{\sf \Omega}_{\bf p}$.
It
will prove convenient to cast  the matrix of vacuum
oscillation frequencies in its traceless form
\begin{equation}
{\sf \Omega}_\omega = -\frac{\omega}{\sqrt{3}} \,\ \lambda_8
- \alpha \frac{\omega}{2} \lambda_3
\end{equation}
where $\lambda_3$ and $\lambda_8$ are the  two diagonal  Gell-Mann matrices,
which read respectively~\cite{Dasgupta:2007ws}
\begin{eqnarray}
\lambda_3 &=& \textrm{diag}(1,-1,0) \nonumber \,\ , \\
\lambda_8 &=& \frac{1}{\sqrt{3}} \textrm{diag}(1,1,-2) \nonumber \,\ . 
\end{eqnarray}
The vacuum oscillation frequency
\begin{equation}
\omega = \frac{\Delta m^2_{\rm atm}}{2E}
\end{equation}
is  associated to the atmospheric mass-square difference
 ${\Delta m^2_{\rm atm}} = m_3^2-m_{1,2}^2$.
The mass hierarchy parameter is 
\begin{equation}
\alpha \equiv \textrm{sgn}(\Delta m^2_{\rm atm}) \frac{\Delta m^2_{\rm sol}}{{\Delta m^2_{\rm atm}}} \,\ ,
\end{equation}
${\Delta m^2_{\rm sol}} = m_2^2-m_{1}^2$ being the solar mass-square difference.
${\Delta m^2_{\rm atm}}>0$
defines
the normal mass hierarchy (NH), while ${\Delta m^2_{\rm atm}}<0$  the inverted hierarchy
 (IH). 
For the numerical illustrations, we take the neutrino mass-squared differences to be $|\Delta m^2_{\rm atm}|= 2\times 10^{-3}$~eV$^2$ and $\Delta m^2_{\rm sol}= 8\times 10^{-5}$~eV$^2$, close to their current best-fit values~\cite{GonzalezGarcia:2010er}.
In SN neutrino flavor conversions, the parameters $(\Delta m^2_{\rm atm},\theta_{13})$ are responsible of conversions
 between $e-y$ states, while $(\Delta m^2_{\rm sol},\theta_{12})$ determine conversions between 
  $e-x$ states~\cite{Dasgupta:2007ws}. The mass hierarchy implies the dominance of $e-y$ conversion effects over 
  the $e-x$ ones. However, are   possible  situations in which one finds an interesting interplay
 between the ``atmospheric'' and ``solar'' sectors~\cite{Friedland:2010sc,Dasgupta:2010cd}, as we will show in the following.

The
matter effect in Eq.~(\ref{eq:Ham}), due to the background electron density $n_e$,
 in the weak interaction basis $(\nu_e,\nu_\mu,\nu_\tau)$ is represented by~\cite{Matt}
\begin{equation}
{\sf V}_{\rm MSW}=\sqrt{2}G_F n_e \textrm{diag}(1,0,0) \,\ .
\end{equation}
Except at very early times
$(t \lesssim 300$~ms) when the effective electron density $n_e$ would
become larger than the neutrino density $n_\nu$ suppressing
the self-induced flavor conversions~\cite{EstebanPretel:2008ni}, one can  
 account for matter effects in the region of collective oscillations just by  choosing small 
(matter suppressed) mixing angles~\cite{Duan:2005cp,EstebanPretel:2008ni}, which we take
to be  $\theta_{13}=\theta_{12}= 10^{-3}$. 
Matter effects in the region of collective oscillations 
(up to a few 100 km) also slightly modify the neutrino
mass-square differences.
 Therefore, we take the effective mass-square differences 
$\Delta {\tilde m}^2_{\rm atm} = \Delta m^2_{\rm atm} \cos \theta_{13}\simeq \Delta m^2_{\rm atm}$
and $\Delta {\tilde m}^2_{\rm sol} = \Delta m^2_{\rm sol} \cos \theta_{12} \simeq 0.4 \Delta m^2_{\rm sol}$%
~\cite{Duan:2008za,EstebanPretel:2008ni}.
  Mikheyev-Smirnov-Wolfenstein (MSW) conversions  typically occur  after collective effects have ceased~\cite{Fogli:2007bk, Dasgupta:2007ws}. Their effects then factorize and can be included separately~\cite{Dighe:1999bi}. Therefore, we neglect them in the following.

Finally, the effective potential due to the neutrino-neutrino
interactions is given by~\cite{Pantaleone:1992eq,Sigl:1992fn,Pantaleone:1992xh,McKellar:1992ja}
\begin{equation}
{\sf V}_{\nu \nu} = \sqrt{2} G_F\int \frac{\textrm{d}^3 {\bf q}}{(2\pi)^3}(\rho_{\bf q}- 
  {\bar\rho}_{\bf q})
(1-{\bf v}_{\bf p}\cdot {\bf v}_{\bf q}) \,\ ,
\end{equation}
where the factor $(1-{\bf v}_{\bf p}\cdot {\bf v}_{\bf q})$ implies \emph{multi-angle} effects for neutrinos
moving on different trajectories~\cite{Duan:2006an}, as explained in Sec.~I.

We will solve the EoM's in the multi-angle case and compare the behavior of the flavor evolution with the 
solution obtained in the \emph{single-angle} approximation~\cite{Duan:2006an}. This  latter requires the occurrence 
of the  self-maintained coherence of the neutrino ensemble, i.e.
that at a given location all the neutrino modes
are aligned with each other, assuming they were aligned
at the source~\cite{Dasgupta:2008cu}.
In this case one obtains  as angle-averaged EoM for the different energy modes, classified in terms of the frequency
$\omega$ 
\begin{equation}
\textrm{i} \partial_r \rho_\omega = [{\sf H}_{{\omega}}, \rho_{\omega}] \,\ ,
\label{eomSA}
\end{equation}
where we have defined the ``reduced neutrino density matrix'' as~\cite{Duan:2008za}
\begin{equation}
\rho_\omega \sim \left\{ 
\begin{array}{cc}
+ \rho_{{\bf q}} & \textrm{if} \,\ \omega >0 \nonumber \\
- {\bar\rho}_{{\bf q}} & \textrm{if} \,\ \omega <0 \,\ , \nonumber 
\end{array}
\right. 
\end{equation}
whose  diagonal components 
\begin{equation}
\rho_{\omega(\alpha \alpha)} (r) = \frac{F_{\nu_\alpha}(\omega,r)}{F_\nu(\omega,r)} \,\ ,
\end{equation}
are normalized to the sum of the fluxes  of all the neutrino species $F_\nu(\omega,r)
= F_{\nu_e}(E,r)+F_{\nu_x}(E,r)+F_{\nu_y}(\omega,r)$ (and analogously for antineutrinos). 
We clarify that when we use the $\omega$-variable, with the notation
$F_\nu(\omega)$ we would mean $F_\nu(E)\times dE/d\omega$. 

Neglecting the matter effects, the single-angle Hamiltonian reads~\cite{Dasgupta:2008cu} 
\begin{equation}
{\sf H}_{\omega} = {\sf \Omega}_{{\omega}_r} + \mu^{\ast}_r \rho \,\ .
\end{equation}
In this equation the vacuum oscillation frequency $\omega$ when projected over the radial direction
becomes~\cite{Dasgupta:2008cu}
\begin{equation}
\omega_r = \frac{\omega}{\langle v_{r}\rangle} \,\ , 
\end{equation}
where
\begin{equation}
\langle v_{r}\rangle = \frac{1}{2}\left[1 +\sqrt{1-\left(\frac{R}{r}\right)^2} \right]
\end{equation}
is the angle-averaged radial neutrino velocity.
The modification  of the vacuum oscillation frequencies is relevant only near the
neutrinosphere, therefore in the following we  neglect it in our estimations.

The neutrino-neutrino interaction term depends on the matrix of total density
\begin{equation}
\rho = \frac{1}{\Phi^0_\nu + \Phi^0_{\overline\nu}} \int_{-\infty}^{+\infty} d\omega F_\nu \rho_\omega \,\,
\end{equation}
normalized to  the sum of the total neutrinos and antineutrinos flux at  the neutrinosphere.

 The radial dependence of the neutrino-neutrino
interaction strength can be written as~\cite{Fogli:2007bk,Dasgupta:2008cu}
\begin{equation}
\mu^\ast_r  = \mu_R \frac{R^2}{2 r^2} C_r \,\ ,
\end{equation}
where 
\begin{equation}
\mu_R = 2 \pi \sqrt{2}G_F (\Phi_{\nu}^0+\Phi_{\bar\nu}^0)  \,\ ,
\end{equation}

represents the strength of the neutrino-neutrino potential at the neutrinosphere.
The $r^{-2}$ scaling comes from the geometrical flux dilution,
and the collinearity factor
\begin{eqnarray}
C_r &=& 4 \left[\frac{1-\sqrt{1-(R/r)^2}}{(R/r)^2}\right]^2 -1  \nonumber \\
& \approx  & \frac{1}{2}\left(\frac{R}{r}\right)^2 \,\ \,\ \,\ \,\ \textrm{for} \,\  r\to\infty \,\ ,
\end{eqnarray}
arises from the $(1-\cos\theta)$ structure of the neutrino-neutrino interaction.
The asymptotic behavior for large $r$ agrees with what one obtains by considering that all neutrinos are launched at $45^{\circ}$ to the 
radial-direction~\cite{EstebanPretel:2007ec}.
The known decline of the neutrino-neutrino interaction strength, $\mu^\ast_r \sim r^{-4}$ 
for $r\gg R$, is evident~\cite{Pantaleone:1992eq}.

The behavior of the neutrino ensemble at large densities in the single-angle case  can be characterized in terms of the invariants of the system. 
Namely, two \emph{lepton-numbers}, which for small in-medium mixing angles   read in flavor basis~\cite{Duan:2008za,Dasgupta:2008cd}
\begin{eqnarray}
{\cal L}_3 &=&  \frac{1}{\Phi^0_\nu + \Phi^0_{\bar\nu}} 
\int_{-\infty}^{+\infty} d\omega F_\nu \textrm{Tr}(\rho_\omega \lambda_3)\nonumber \\
{\cal L}_8 &=&  \frac{1}{\Phi^0_\nu + \Phi^0_{\bar\nu}} 
\int_{-\infty}^{+\infty} d\omega F_\nu \textrm{Tr}(\rho_\omega \lambda_8) \,\ ,  \nonumber \\
\label{eq:lepton}
 \end{eqnarray}
and the \emph{effective energy} of the system~\cite{Raffelt:2007yz}
\begin{equation}
{\cal E} = -\frac{1}{\sqrt{3}}\textrm{Tr}(\tilde{\rho}\lambda_8) -
 \frac{\alpha}{2} \textrm{Tr}(\tilde{\rho}\lambda_3) +\frac{\mu^\ast_r}{2}
  \textrm{Tr}(\rho^2) \,\ ,
  \label{eq:energy}
\end{equation}
where
\begin{equation}
\tilde{\rho}  = \frac{1}{\Phi^0_\nu + \Phi^0_{\bar\nu}}
\int_{-\infty}^{+\infty} d\omega \,\ \omega \,\   F_\nu \rho_\omega \,\ , 
\end{equation}
is the first momentum of the density matrix.

\section{Multi-angle effects for different neutrino fluxes }

In this Section we compare the results of the SN neutrino flavor evolution, obtained using the 
 single-angle [Eq.~(\ref{eomSA})] and multi-angle [Eq.~(\ref{eomMA})] EoM's for the three representative SN flux models
 introduced in Eq.~(\ref{eq:cases}). 
 For the sake of  brevity,  we will show only results of the supernova neutrino flavor evolution
in the case of inverted neutrino mass hierarchy. We think that this case is  more interesting, since possible three-flavor
 effects, associated with $\Delta m^2_{\rm sol}$ have been recently found in the single-angle scheme for SN neutrino fluxes
with multiple crossing points~\cite{Friedland:2010sc,Dasgupta:2010cd}.  We will show that multi-angle effects can strongly suppress the impact of these three-flavor conversions.  

In our numerical simulations we fix the value of the neutrino-neutrino interaction strength at the 
 neutrinosphere
 \begin{equation}
 \mu^\ast_r (R) = 2.1 \times 10^{6} \,\ \textrm{km}^{-1} \,\ ,
 \end{equation}
 unless otherwise stated. This choice will imply that neutrino luminosities in the three
 different cases will be (in units of $10^{51}$~erg/s)
 \begin{eqnarray}
 & & L_{\nu_e}= 2.40 \,\ , \,\   L_{{\bar\nu}_e}= 2.00 \,\ , \,\  L_{\nu_x}= 1.50 \,\ , \nonumber \\
 & & L_{\nu_e}= 1.20 \,\ , \,\   L_{{\bar\nu}_e}= 1.34 \,\ , \,\  L_{\nu_x}= 2.14 \,\ , \nonumber \\
 & & L_{\nu_e}= 1.16 \,\ , \,\,  L_{{\bar\nu}_e}= 1.41 \,\ , \,\  L_{\nu_x}= 2.14 \,\ . \nonumber \\
 \end{eqnarray}
In order to get stable results in our multi-angle simulations we  typically use $\sim1500$ angular modes.
Finally, to saturate self-induced oscillation effects, we integrate the EoM's till $r=10^3$~km.
Technical
details are discussed in the Appendix: in the following we focus only on the results and their
interpretation.

\begin{figure}[!t]
\begin{center}  
\includegraphics[width=\columnwidth]{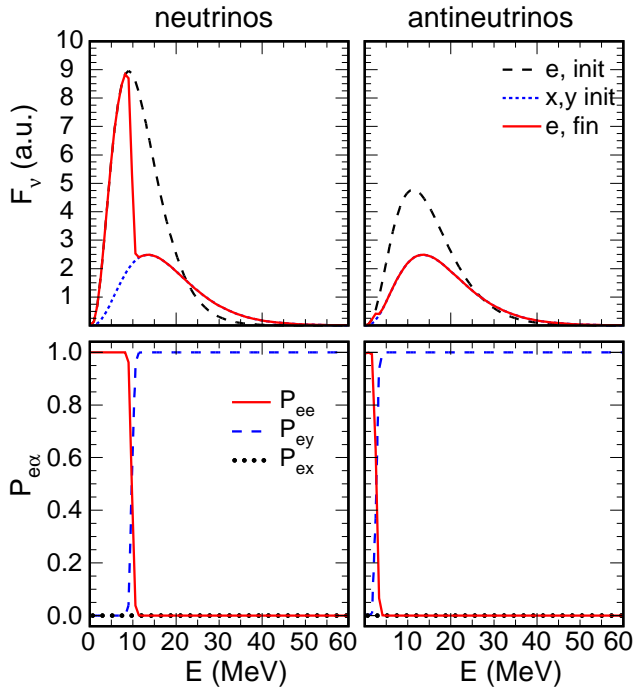}  
\end{center}  
\caption{Case with  
$\Phi^0_{\nu_e}: \Phi^0_{{\bar{\nu}}_e}:\Phi^0_{\nu_x} = 2.40:1.60:1.0$. Three-flavor evolution in
 inverted mass hierarchy for the 
\emph{single-angle} case
for neutrinos (left panels) and antineutrinos (right panels). Upper panels:
Initial energy spectra  for $\nu_e$ (long-dashed curve) and $\nu_{x,y}$
(short-dashed curve)
 and for   $\nu_e$ after collective oscillations (solid curve).
Lower panels: probabilities $P_{ee}$ (solid
red curve), $P_{ey}$ (dashed blue curve), $P_{ey}$ (dotted black curve).
\label{fig:1}}  
\end{figure}  

\begin{figure}[!t]
\begin{center}  
\includegraphics[width=\columnwidth]{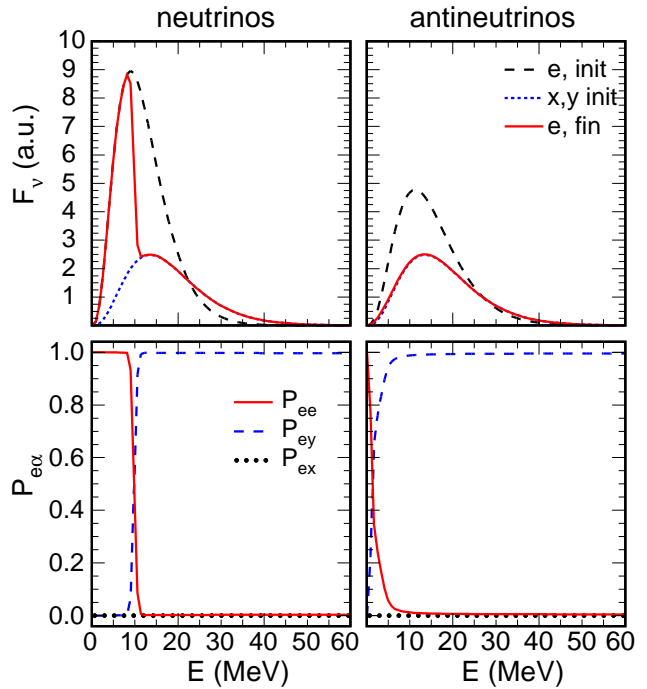}  
\end{center}  
\caption{As in Fig.~1, but for the
\emph{multi-angle} case.
\label{fig:2}}  
\end{figure}  

\subsection{Spectrum with a single crossing }

We start our investigation with  the SN neutrino flux ordering
$\Phi^0_{\nu_e}: \Phi^0_{{\bar{\nu}}_e}:\Phi^0_{\nu_x} = 2.40:1.60:1.0$,
as representative  of the accretion phase. 
This case has been the  benchmark for most of the previous multi-angle  studies (see, e.g.,~\cite{Duan:2006an,Fogli:2007bk}). 
As known, in this case the dynamics can be studied into the $e-y$  two-neutrino system associated with $(\Delta m^2_{\rm atm}, \theta_{13})$. Three-flavor
effects in the $e-x$ sector, associated with ($\Delta m^2_{\rm sol}, \theta_{12}$) are negligible~\cite{Dasgupta:2007ws,Fogli:2008fj}.

In Fig.~1 we represent the initial neutrino  fluxes  at the neutrinosphere for all the different species and 
the final electron (anti)neutrino fluxes after collective oscillations (at $r= 10^3$~km) in the single-angle approximation (upper panels) and 
 the corresponding
$P_{ee} = P(\nu_e\to\nu_e)$, $P_{ex} = P(\nu_e\to\nu_x)$ and $P_{ey} = P(\nu_e\to\nu_y)$ conversion probabilities (lower panels). 
Neutrinos are shown in the left panels and antineutrinos in the right panels.  
The corresponging results for the multi-angle case are given in Fig.~2.
We stress  that, except in the high-energy tails,  the original neutrino spectra in the case under study always have an excess of electron (anti)neutrinos
over the non-electron species, i.e. $F_{\nu_e} > F_{\nu_x}$ and $F_{{\overline\nu}_e} > F_{{\overline\nu}_x}$.
In the
frequency  variable $-\infty <\omega<+\infty$, the crossing point at $E\to \infty$ ($\omega=0$) and the other two at
$E\gtrsim 20$~MeV appear so narrowly
 spaced, that the neutrino spectra  superficially appear with only a \emph{single crossing} point at $\omega=0$, where  $F_{\nu_e} = F_{\nu_x}=0$ and  $F_{{\overline\nu}_e} = F_{{\overline\nu}_x}=0$. We address the interested reader to 
Fig.~3 of~\cite{Dasgupta:2009mg} and to the relative discussion  on the case   neutrino spectrum
effectly behaving as a single-crossing one. We will show how this property is
crucial
in characterizing the flavor evolution.

Concerning the neutrino fluxes after self-induced oscillations, in the single-angle case one finds a swap between $\nu_e$ and $\nu_y$ spectra above $E\simeq 10$~MeV, producing the typical split in the final neutrino spectra.
 For antineutrinos, the swap between ${\bar\nu}_e$ and ${\bar\nu}_y$ is almost complete, the splitting energy
$E\simeq 2$~MeV being very low. 
Neglecting this low-energy feature~\cite{Fogli:2008pt},   the position of the $\nu_e$ split can be calculated using the conservation of the lepton number ${\cal L}_8$ in Eq.~(\ref{eq:lepton}) (see, e.g.,~\cite{Raffelt:2007cb,Raffelt:2007xt,Duan:2007bt}).
In the multi-angle case, the swap features remain unchanged, except for the smearing
of the  low-energy anti-neutrino spectral split,  shown also in~\cite{Fogli:2008pt}.

In Fig.~3 and 4 we represent the radial evolution of the diagonal elements of the density matrix $\rho_{ee}$,  $\rho_{yy}$, $\rho_{xx}$ for different energy modes for neutrinos (left panels) and antineutrinos (right panels).
In particular, in the $\rho_{ee}$ panels the order of the energy modes is $E= 0.91, 5, 11, 43$~MeV
going from the curve starting with the highest value  to the lowest one. This order is reversed in the $\rho_{yy}$ and $\rho_{xx}$ panels. 
 Figure~3 represents the single-angle evolution, while Fig.~4 is for the multi-angle evolution, where the density matrix elements have been integrated over the angular distribution.  Except for very low-energy 
antineutrino modes, we find that the evolution of the density matrix is rather similar in the single-angle and multi-angle case. We find the presence of synchronized oscillations~\cite{Kostelecky:1993dm,Hannestad:2006nj} till $r \simeq 85$~km. Till there  all the $\rho_{\omega}$ stay pinned to their original value. 

\begin{figure}[!t]
\begin{center}  
\includegraphics[width=\columnwidth]{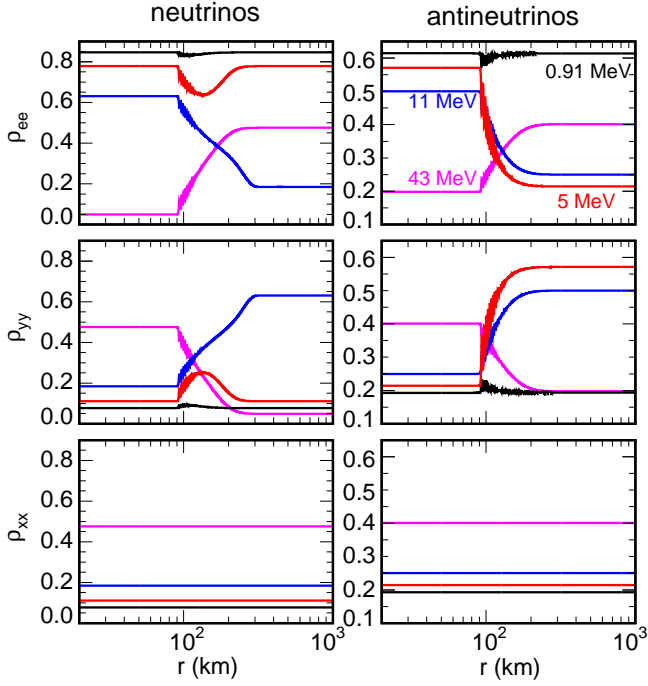}  
\end{center}  
\caption{Case with $\Phi^0_{\nu_e}: \Phi^0_{{\bar{\nu}}_e}:\Phi^0_{\nu_x} = 2.40:1.60:1.0$. Three-flavor evolution in inverted mass hierarchy in  the \emph{single-angle} case.
Radial evolution of the diagonal components of the density matrix $\rho$ for neutrinos (left panels) and
antineutrinos (right panels)  for  different energy modes.
\label{fig:3}} 
\end{figure}  

\begin{figure}[!t]
\begin{center}  
\includegraphics[width=\columnwidth]{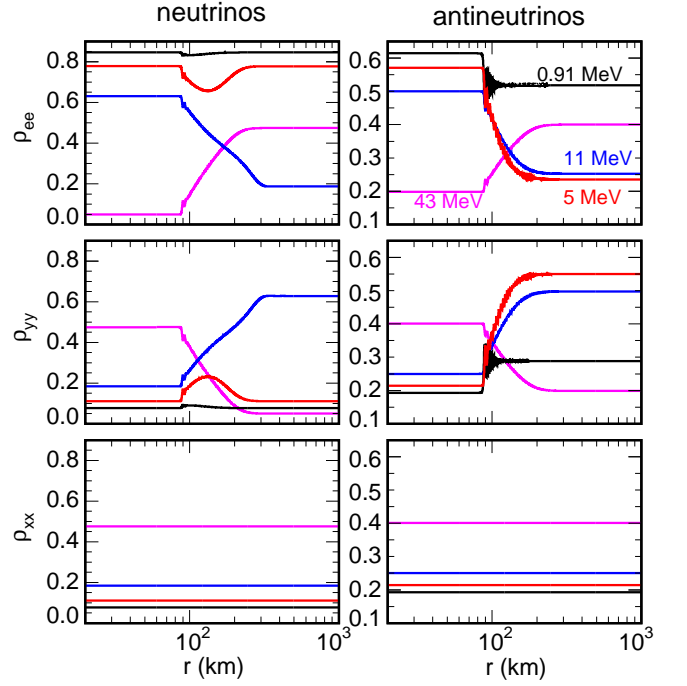}  
\end{center}  
\caption{As in Fig.~3, but for the \emph{multi-angle} case. 
\label{fig:4}} 
\end{figure}  

Significant flavor conversions  start  only after synchronization  when bipolar oscillations~\cite{Hannestad:2006nj}  start to swap  the flavor content of the system.
The synchronization radius can be found through the condition~\cite{Fogli:2007bk}
\begin{equation}
\mu^\ast_r > \frac{4 \,\ \omega_{\rm s}}{\sqrt{3}}  \frac{\textrm{Tr}(\sigma\lambda_8)}{[\textrm{Tr}(\rho \lambda_8)]^2} \,\  .
 \label{eq:synccond}
\end{equation}
Using the alignment approximation
for the two blocks of neutrinos and antineutrinos, the synchronization frequency for $e-y$ conversions,  reads~\cite{Fogli:2007bk}
\begin{eqnarray}
\omega_{\rm s} &=& \frac{\int_{0}^{+\infty} d{\omega} \,\ \omega (F^0_{\nu_e}(E)-F^0_{\nu_y}(\omega))}
{2 (\Phi^0_{\nu_e}-\Phi^0_{{\nu}_y})} \nonumber \\
&+& \frac{\int_{0}^{+\infty} d{\omega} \,\ \omega (F^0_{{\bar\nu}_e}(\omega)-F^0_{{\bar\nu}_y}(\omega))}
{2 (\Phi^0_{{\bar\nu}_e}-\Phi^0_{{\bar\nu}_y})} \,\ \nonumber \\
&=& \frac{2\Delta m^2_{\rm atm}}{3}  \bigg[\frac{ 
L_{\nu_e}/\langle E_{\nu_{e}}\rangle^2 - L_{\nu_x} /\langle E_{\nu_{x}}\rangle^2}
{L_{\nu_e}/\langle E_{\nu_{e}}\rangle - L_{\nu_x} /\langle E_{\nu_{x}}\rangle} \,\  \nonumber \\
&+& \frac{ 
L_{\overline\nu_e}/\langle E_{\overline\nu_{e}}\rangle^2 - L_{\nu_x} /\langle E_{\nu_{x}}\rangle^2}
{L_{\overline\nu_e}/\langle E_{\overline\nu_{e}}\rangle - L_{\nu_x} /\langle E_{\nu_{x}}\rangle}
\bigg] \,\ \nonumber  \\
&\simeq& 0.68 \,\   \textrm{km}^{-1} \,\ , \nonumber 
\end{eqnarray}
for our input fluxes. The function
\begin{displaymath}
\sigma =  \frac{1}{\Phi^0_\nu + \Phi^0_{\overline\nu}} \int_{-\infty}^{+\infty} d\omega \,\ 
\textrm{sgn}(\omega)
 F_\nu {\rho}_{\omega}  \,\ . 
 \nonumber
\end{displaymath}
Then 
\begin{eqnarray}
& & \textrm{Tr}(\sigma\lambda_8) = \frac{\rho_{ee} +{\bar\rho}_{ee}- 2 \rho_{xx}}{\sqrt{3}} = \nonumber \\
&=& \frac{1}{\sqrt{3}}\frac{L_{\nu_e}\langle E_{\overline\nu_{e}}
\rangle \langle E_{\nu_x} \rangle + L_{\overline\nu_e}   \langle E_{\nu_{e}} \rangle \langle E_{\nu_x} \rangle
-2  L_{\nu_x}  \langle E_{\overline\nu_{e}} \rangle \langle E_{\nu_{e}} \rangle}
{L_{\nu_e}    
\langle E_{\nu_x} \rangle  \langle E_{\overline\nu_{e}} \rangle + L_{\overline\nu_e}  \langle E_{\nu_{e}}\rangle \langle E_{\nu_x} \rangle 
+ 4 L_{\nu_x}  \langle E_{\nu_{e}}\rangle \langle E_{\overline\nu_{e}} \rangle} \,\
\nonumber \\
&\simeq& 0.144 \,\ ,  
\nonumber 
\end{eqnarray}
and 
\begin{eqnarray}
& & \textrm{Tr}(\rho\lambda_8) = 
\frac{\rho_{ee}-\bar\rho_{ee}}{\sqrt{3}} = \nonumber \\
&=& \frac{1}{\sqrt{3}} \frac{\langle E_{\nu_x} \rangle [L_{\nu_e} \langle E_{\overline\nu_{e}}
\rangle - L_{\overline\nu_e} \langle E_{\nu_{e}}\rangle ]}{L_{\nu_e}    
\langle E_{\nu_x} \rangle  \langle E_{\overline\nu_{e}} \rangle + L_{\overline\nu_e}  \langle E_{\nu_{e}}\rangle \langle E_{\nu_x} \rangle 
+ 4 L_{\nu_x}  \langle E_{\nu_{e}}\rangle \langle E_{\overline\nu_{e}} \rangle} \,\ \nonumber  \\
&\simeq& 0.06 \,\ .  \nonumber 
\end{eqnarray} 
These numbers  imply that bipolar conversions would start when $\mu^{\ast}_r \simeq 100 \,\ \omega_H$, i.e.
at $r \simeq 85$~km, as observed in the Fig.~3. 

We stress that the presence of synchronization at low-radii in this case is crucially related with the original spectrum with a single crossing  point.
As discussed in~\cite{Raffelt:2008hr},  energy conservation at large neutrino densities (i.e. when $\mu^\ast_r \to \infty$)  implies  that  $\rho^2$ 
and thus $\rho$ are conserved [see Eq.~(\ref{eq:energy})],
and therefore behave as   collective objects.
If $\rho$ is conserved, energy conservation implies that also    $\textrm{Tr}(\tilde{\rho}\lambda_8)$
has to be conserved (neglecting the subleading three-flavor effects associated to $\Delta m^2_{\rm sol})$. 
For a single-crossed spectrum, the quantity $\textrm{Tr}(\tilde{\rho}\lambda_8) =
 [(\tilde{\rho}_{ee}-\tilde{\rho}_{yy})+ (\tilde{\bar\rho}_{ee}-\tilde{\rho}_{yy})]/\sqrt{3}$ is maximal~\cite{Dasgupta:2009mg}.
 Therefore moving any energy mode $\rho_\omega$ from its initial value would make the energy of the system larger.  In this situation, each  $\rho_\omega$ must remain pinned 
with the global density matrix $\rho$. Therefore,   only the synchronization among different modes is possible~\cite{Raffelt:2008hr}.
Since during the synchronization phase the different $\rho_\omega$   remain aligned to their original  value, multi-angle affects are suppressed.

Then, during the phase of the  spectral swapping, the spectrum
near the crossing point acts like an inverted pendulum~\cite{Dasgupta:2009mg}. The swap sweeps through the spectrum on each
side of the crossing, and the modes at the edge of the
swap precess at an average oscillation frequency ${\kappa}\simeq \Delta m^2_{\rm atm}/2E \simeq 0.5$~km$^{-1}$ for a typical energy $E \simeq   10$~MeV in the region of the swap. Since, the
length scale $l_{\mu} \equiv  |d \ln \mu^{\ast}_r(r)/dr|^{-1} = r/4$ is larger than ${\kappa}^{-1}$, the evolution is adiabatic
concerning the swapping dynamics~\cite{Friedland:2010sc,Dasgupta:2010cd}.
However, the time-scale of  multi-angle effects is determined by the bipolar oscillation frequency
 $\omega_{\rm H} \sim \sqrt{2 \omega \mu^{\ast}_r} \sim r^{-2}$~\cite{Raffelt:2007yz} that  decreases faster than $l_{\mu}^{-1}$. 
Qualitatively,
the  effect of the synchronization is to postpone conversions  at large radii, where
multi-angle effects would require more ``time'' to develop since $\mu^{\ast}_r$ is smaller,
and  the relatively fast
decrease of $\mu^{\ast}_r$ would reduce the adiabaticity   for multi-angle effects. As a consequence  these do not grow significantly.  
This is qualitatively the origin of the  so called ``quasi single-angle'' behavior~\cite{EstebanPretel:2007ec}, observed for neutrino flux ordering typical of the accretion phase.
We have explicitly checked that modifying artificially the adiabaticity of the evolution, i.e. choosing a very
slow decrease of $\mu^{\ast}_r$ multi-angle decoherence is unavoidable also in the 
case we are considering. Therefore, as suggested in~\cite{EstebanPretel:2007ec}, the absence of multi-angle 
decoherence during the accretion phase seems due to a luckily conspiracy of different
time-scales.   
Unfortunately, till now it has not been  developed yet a complete theory
for that. Therefore, the interpretation
of these effects is mostly based on  the experience gained through numerical observations.

\subsection{Spectrum with multiple crossings}

We pass now to the case
 $\Phi^0_{\nu_e}:\Phi^0_{{\bar{\nu}}_e}:\Phi^0_{\nu_x} =0.85:0.75:1.0$  representative of recent simulation results for
the neutrino flux ordering during the cooling phase~\cite{Raffelt:2003en}.  
This case has been widely studied in the single-angle approximation in~\cite{Dasgupta:2009mg,Friedland:2010sc,Dasgupta:2010cd} and
corresponds to spectra that in $\omega$ variable present well separated   multiple crossings, in which multiple spectral splits can arise 
around the crossing points (see Fig.~2 of~\cite{Dasgupta:2009mg}).
 Moreover, in this case peculiar three-flavor effects,  associated with 
conversions between $e$ and $x$ states have been recently discussed~\cite{Friedland:2010sc,Dasgupta:2010cd}.
Multi-angle simulations in this case have shown a smearing of the splitting features~\cite{Dasgupta:2009mg,Duan:2010bf}
and a suppression the flavor evolution
at low-radii~\cite{Duan:2010bf,animations}.

In Fig.~5 we show the initial  (anti)neutrino  spectra for all the flavors and the final electron (anti)neutrino ones for the single-angle case (upper panels)
and the corresponding conversion probablities at the end of the flavor evolution.
Multi-angle results are shown in Fig.~6.
Starting with the single-angle case, we see that both $\nu_e \leftrightarrow \nu_y$ as well as $\bar\nu_e \leftrightarrow \bar\nu_y$ swaps appear at intermediate energies $5$ MeV $\lesssim E \lesssim 20$ MeV. Moreover,  for $E \gtrsim 25$ MeV, there are additional $\nu_e \leftrightarrow \nu_x$ and $\bar\nu_e \leftrightarrow \bar\nu_x$ swaps.
As result of this complex dynamics, in the  single-angle case the oscillated $\nu_e$ spectrum shows
a single split, the one at low-energies, producing 
the swap with $\nu_y$ after the split, while the high-energy split is canceled
by the further swap between  $\nu_e$ and $\nu_x$.
We observe that in the antineutrino sector the spectral swaps are not complete, due to
smaller spectral differences between ${\bar\nu}_e$ 
and ${\bar\nu}_x$ that lead to
a less adiabatic evolution, as explained in~\cite{Dasgupta:2010cd}.

Passing  to the multi-angle case,   the differences  with the single-angle case in the final neutrino spectra and 
in the conversion probabilities  are significant. In general, all the splitting features are 
less pronounced than in the single-angle case.
 Starting with the neutrino case, we see that the low-energy spectral split is 
smeared. More remarkably, the high-energy swap between $\nu_e$ and $\nu_x$ is strongly suppressed. As a consequence,  the final neutrino spectra keep track only of the swap between $\nu_e$ and $\nu_y$, which now is less sharp than in the single-angle case. This effect would produce a high-energy splitting feature in the final     $\nu_e$ spectrum at  $E\simeq 25$~MeV,
not present in the single-angle three-flavor calculations. For antineutrinos, the impact of multi-angle effects is even stronger:
  flavor transformations are
significantly suppressed, except   in the energy range close to crossing point  of the spectra, where two-flavor
 ${\bar\nu}_e \to {\bar\nu}_y$ transformations take place.

\begin{figure}[!t]
\begin{center}  
\includegraphics[width=\columnwidth]{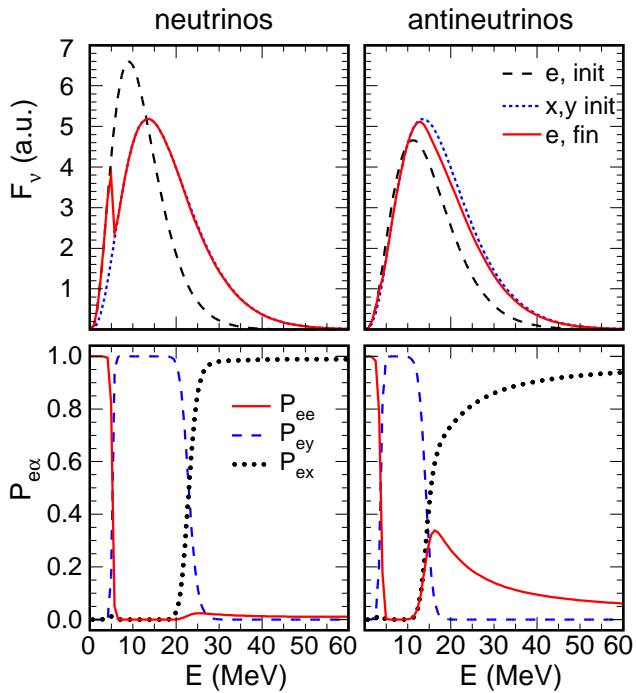}  
\end{center}  
\caption{Case with  $\Phi^0_{\nu_e}:\Phi^0_{{\bar{\nu}}_e}:\Phi^0_{\nu_x} =0.85:0.75:1.0$.
Three-flavor evolution in the \emph{single-angle} case
for neutrinos (left panels) and antineutrinos (right panels). Upper panels:
Initial energy spectra  for $\nu_e$ (long-dashed curve) and $\nu_{x,y}$
(short-dashed curve)
 and for   $\nu_e$ after collective oscillations   (solid curve).
Lower panels: probabilities $P_{ee}$ (solid
red curve), $P_{ey}$ (dashed blue curve), $P_{ey}$ (dotted black curve).
\label{fig:5}}  
\end{figure}  

\begin{figure}[!t]
\begin{center}  
\includegraphics[width=\columnwidth]{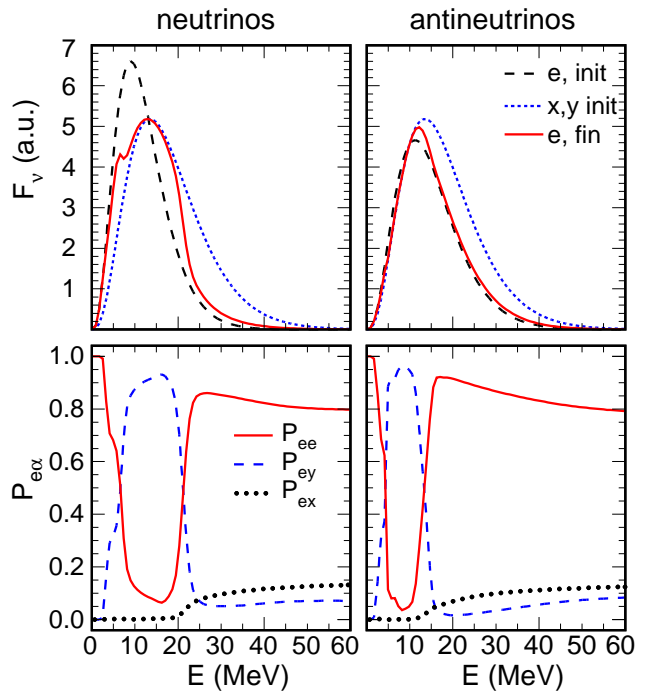}  
\end{center}  
\caption{As in Fig.~5, but for the
\emph{multi-angle} case.
\label{fig:6}}  
\end{figure}  

 To explain the above observations, we represent the radial evolution of the diagonal elements of the density matrix $\rho_{ee}$,
 $\rho_{yy}$, $\rho_{xx}$ for different energy modes,
 for neutrinos (left panels) and antineutrinos (right panels) in  Figs.~7 and 8.
Figure~7 represents the single-angle case, while Fig.~8 represents the multi-angle case.
In this latter case, the density matrix  elements have been  integrated over the angular distribution.
In the $\rho_{ee}$ panels the order of the energy modes is $E= 3.3, 5.7, 19, 30$~MeV
going from the  curve starting with the highest value to the lowest one. This order is reversed in the $\rho_{yy}$ and $\rho_{xx}$ panels.   
 Starting with the single-angle case, we see  that differently from 
the case studied in the Sec.~III A,  flavor conversions are possible at low radii ($r\gtrsim 30$~km) in a region where we would have expected synchronization.
 The effect of the small in-medium mixing is only to logarithmically delay the onset of the flavor conversions.

As explained in~\cite{Raffelt:2008hr}, the crucial difference with respect to the previous case is that, applying the energy conservation [Eq.~(\ref{eq:energy})] to the case of multiple-crossed spectra, it is possible to conserve $\textrm{Tr}(\tilde{\rho}\lambda_8)$
and $\alpha \textrm{Tr}(\tilde{\rho}\lambda_3)$, by flipping parts of the original neutrino spectra
around the crossing points. Therefore, the synchronization behavior found in the case of single-crossed spectra is not necessarily
stable in this case. Indeed,  in~\cite{Raffelt:2008hr} it has been discussed the possibility
of a novel form of flavor conversions for large $\mu^\ast_r$ in terms of a self-induced \emph{parametric resonance}
that would destabilize the synchronization. 
The origin of this effect is that since  each $\rho_{\omega}$ would precess around the Hamiltonian ${\sf H}_\omega$ with 
a frequency $\mu^\ast_r$, ${\sf H}_\omega$ itself must vibrate with the same frequency.
In this situation, for large $\mu^{\ast}_r$ the total density matrix  $\rho$ continues to 
behave as a collective object, but is not static and vibrates itself
with a frequency $\mu^\ast_r$. 
For single-crossed spectra, the energy conservation prevents any seizable change in the different $\rho_{\omega}$,
therefore the synchronized behavior prevails. Conversely, in presence of multiple-crossed neutrino spectra,
the different  $\rho_\omega$'s 
do not remain  aligned with the global  ${\rho}$ and 
start to librate relative to each other.

 We also note that in this case three-flavor effects, associated with 
$(\Delta m^2_{\rm sol},\theta_{12})$ are present. In particular, from the figure, one realizes that
collective librations first start in the $e-y$ system (at $r\gtrsim 30$~km), and then trigger the $e-x$ conversions (at $r\gtrsim 60$~km). The explanation of this dynamics has been recently given in~\cite{Dasgupta:2010cd}, where we redirect the interested reader.

Passing to the multi-angle case of  Fig.~8, we find 
that now the low-radii flavor conversions are suppressed.
 This effect    has been recently described in~\cite{Duan:2010bf}, even if
there it has not made the connection between this behavior and the multi-angle suppression of the parametric resonance. 
This effect has been pointed out in~\cite{Raffelt:2008hr} for generic neutrino gases with
 half-isotropic neutrino distributions, in cases  where in the single-angle
scheme the parametric resonance was found.  The point is that in the  multi-angle case,  for large $\mu^\ast_r$ the different $\rho_{\omega}$'s would precess
with different velocities whose spread, due to multi-angle effects,  is much larger than $\omega$. As a consequence of this dispersion, there cannot
exist a collective parametric resonance for all the neutrino modes.
Therefore, the collective
behavior found in the idealized single-angle case  is ``fragile'' and easily suppressed
by multi-angle effects in the realistic anisotropic supernova environment.
In this case, as long as  multi-angle effects are dominant, they  would suppress any flavor conversion at small radii.
As discussed  in~\cite{Duan:2010bf}, the role of multi-angle effects in this case is   analogous to the one that  they have  in presence of    a large matter term
($n_e \gg  n_\nu$)~\cite{EstebanPretel:2008ni}. There, multi-angle effects introduce
a significant dispersion for the matter potential encountered by neutrinos  on different trajectories, that once more would suppress self-induced conversions.

\begin{figure}[!t]
\begin{center}  
\includegraphics[width=\columnwidth]{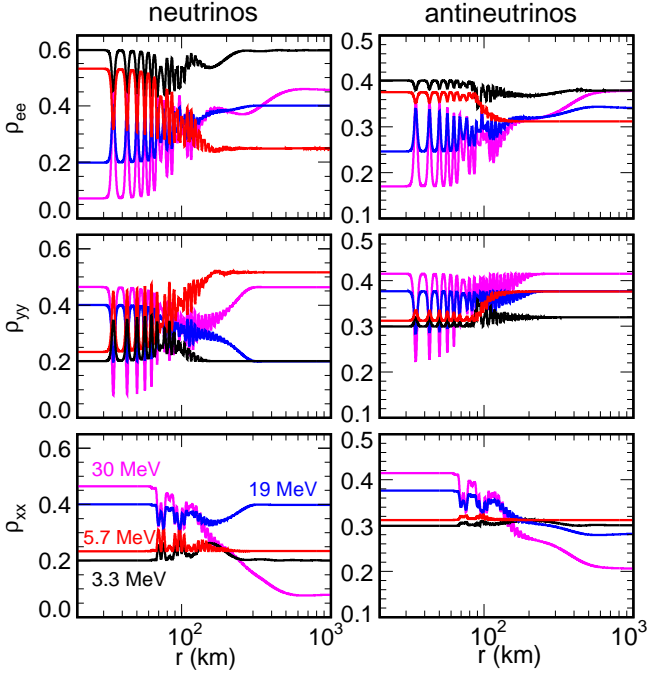}  
\end{center}  
\caption{Case   with  $\Phi^0_{\nu_e}:\Phi^0_{{\bar{\nu}}_e}:\Phi^0_{\nu_x} =0.85:0.75:1.0$.
 Three-flavor evolution in inverted mass hierarchy in single-angle case.
Radial evolution of the diagonal components of the density matrix $\rho$ for neutrinos (left panels) and
antineutrinos (right panels)  for  different energy modes.
\label{fig:7}}  
\end{figure}  

\begin{figure}[!t]
\begin{center}  
\includegraphics[width=\columnwidth]{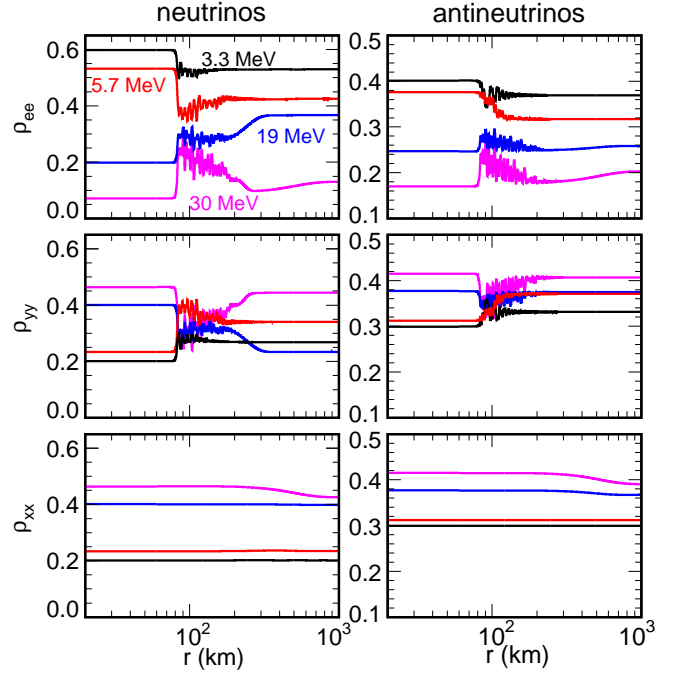}  
\end{center}  
\caption{As in Fig.~7, but for the
\emph{multi-angle} case.
\label{fig:8}} 
\end{figure}  

 In the presence of  multi-angle suppression, 
 we expect that flavor conversions would start only when  neutrino-neutrino interactions are not strong enough to maintain the
 collective behavior  of $\rho$~\cite{Duan:2010bf}.
Neutrino modes would  remain frozen to their initial condition, till
the neutrino-neutrino 
interaction $\mu^\ast_r$ becomes comparable with the averaged vacuum oscillation frequency  
of the neutrino ensemble~\cite{Dasgupta:2007ws}
\begin{equation}
\langle \omega \rangle = \frac{\textrm{Tr}({\tilde\rho} \lambda_8)}{\textrm{Tr}(\rho \lambda_8)}
= \frac{\tilde{\rho}_{ee}+{\tilde{\bar{\rho}}}_{ee}-2\tilde{\rho}_{yy}}{\rho_{ee}-\bar\rho_{ee}} \,\ ,
\end{equation}
where for our input spectra 
\begin{eqnarray}
 & &\tilde{\rho}_{ee}+{\tilde{\bar{\rho}}}_{ee}-2\tilde{\rho}_{yy}  \nonumber \\
 &=&\frac{2 \Delta m^2_{\rm atm}}{3}
 \frac{ L_{\nu_e}/\langle E_{\nu_{e}}\rangle^2 + L_{\overline\nu_e} \langle E_{\overline\nu_{e}}
\rangle^2 -2 L_{\nu_x}/\langle E_{\nu_{x}}\rangle^2
}{L_{\nu_e}/\langle E_{\nu_{e}}\rangle + L_{\overline\nu_e} \langle E_{\overline\nu_{e}}
\rangle + 4  L_{\nu_x}/\langle E_{\nu_{x}}\rangle} \,\ \nonumber  \\
&\simeq& 1.1 \times 10^{-2} \,\ \textrm{km}^{-1} \,\ , \nonumber
\end{eqnarray}
 
and

\begin{eqnarray}
& & \rho_{ee}-{\bar\rho}_{ee}  \nonumber \\
&=& \frac{L_{\nu_e}\langle E_{\overline\nu_{e}}
\rangle \langle E_{\nu_x} \rangle + L_{\overline\nu_e}   \langle E_{\nu_{e}} \rangle \langle E_{\nu_x} \rangle
-2  L_{\nu_x}  \langle E_{\overline\nu_{e}} \rangle \langle E_{\nu_{e}} \rangle}
{L_{\nu_e}    
\langle E_{\nu_x} \rangle  \langle E_{\overline\nu_{e}} \rangle + L_{\overline\nu_e}  \langle E_{\nu_{e}}\rangle \langle E_{\nu_x} \rangle 
+ 4 L_{\nu_x}  \langle E_{\nu_{e}}\rangle \langle E_{\overline\nu_{e}} \rangle} \,\
\nonumber \\
&\simeq & 1.6 \times 10^{-2} \,\ ,
\end{eqnarray}
which numerically  leads to  $\langle \omega \rangle \simeq 0.68$~km$^{-1}$.

We expect flavor conversions to start in the $e-y$ sector  when 
\begin{equation}
\langle \omega \rangle \simeq \sqrt{3} \mu^\ast_r \textrm{Tr}(\rho \lambda_8) \,\ ,
\label{eq:multisup}
\end{equation}
that would correspond to $\mu^\ast_r \simeq 60 \,\  \langle \omega \rangle$.
For our input spectra, we would get as onset radius $r_{\rm ons} \simeq 92$~km,
in qualitative agreement with the  simulation of Fig.~8. We note that numerically the conversions' onset  radius 
in this case is very similar to the one of Sec.~III A, even if the conditions that determine it are different,
as well as  the further flavor evolution.
In particular, we expect that at large radii the evolution in this case would be  less adiabatic than in the case of
Sec.~III A. 
As explained in~\cite{Dasgupta:2009mg}, in the final phases of the swapping dynamics, the
neutrino and antineutrino spectra evolve quite independently,
and the precession frequencies of the two blocks
are not governed by a common $\mu^{\ast}_r$, but by individual $\mu^{\ast}_r$'s
proportional to the flux differences $|F_{\nu_e}-F_{\nu_x}|$ ($|F_{{\bar\nu}_e}-F_{{\bar\nu}_x}|$). They
behave essentially as two uncoupled oscillators because
the neutrino-neutrino interaction $\mu^{\ast}_r$ is now smaller than
the frequency difference of the two blocks.
Since in this case the flux differences are smaller than in the 
previous example of Sec.~III A, the effective precession frequencies are smaller, with a resultant less adiabatic
evolution. This would explain the more pronounced smearing of the  sharp splitting features observed in this case in
the multi-angle simulations. 

\begin{figure}[!t]
\begin{center}  
\includegraphics[width=\columnwidth]{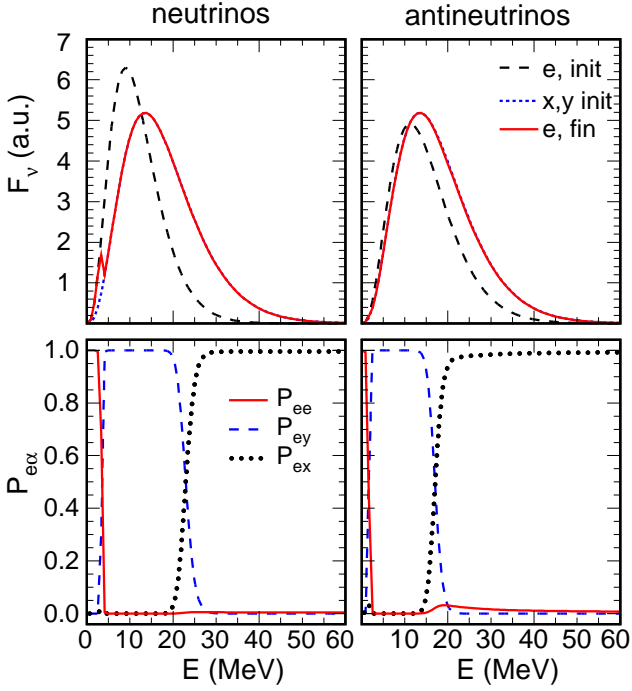}  
\end{center}  
\caption{Case with $\Phi^0_{\nu_e}:\Phi^0_{{\bar{\nu}}_e}:\Phi^0_{\nu_x} =0.81:0.79:1.0$.
Three-flavor evolution in the \emph{single-angle} case
for neutrinos (left panels) and antineutrinos (right panels). Upper panels:
Initial energy spectra  for $\nu_e$ (long-dashed curve) and $\nu_{x,y}$
(short-dashed curve)
 and for   $\nu_e$ after collective oscillations   (solid curve).
Lower panels: probabilities $P_{ee}$ (solid
red curve), $P_{ey}$ (dashed blue curve), $P_{ey}$ (dotted black curve).
\label{fig:9}}  
\end{figure}  

\begin{figure}[!t]
\begin{center}  
\includegraphics[width=\columnwidth]{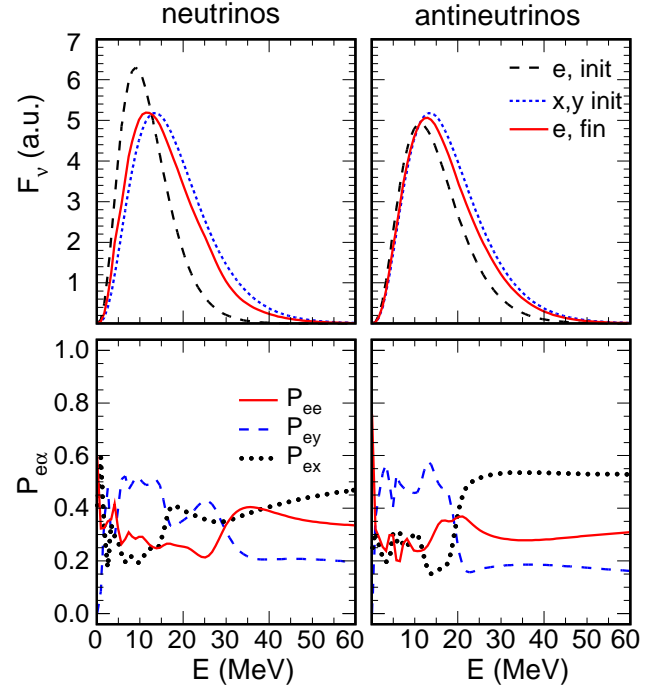}  
\end{center}  
\caption{As in Fig.~9, but for the
\emph{multi-angle} case.
\label{fig:10}}  
  \end{figure}  

The effect of the multi-angle delay is that flavor conversions start in a region in which the 
neutrino-neutrino interaction strength is weaker than in the corresponding single-angle case.
The  typical time-scale at which the off-diagonal components in the density matrix  grow
is given by the bipolar period which,  for
 $e-y$ transitions, is 
$\tau_H \sim  (2\omega \mu^{\ast}_r)^{-1/2}$~\cite{Hannestad:2006nj}. The delay of the conversions means that they start at a  smaller
$\mu^{\ast}_r$ where they require more time to develop.
 Moreover,  going at larger $r$ makes the evolution less adiabatic, resulting in  weaker
flavor conversions than in the single-angle case. This would explain 
the suppression of the $e-x$ conversions. 
 At this regard,
we remind that $e-x$ conversions  are triggered by the dominant
$e-y$ conversions, via the $\theta_{13}$ coupling between the two sectors~\cite{Dasgupta:2010cd}.
Due to the suppression of $e-y$ conversions, the $e-x$ transitions would also be  delayed.
The typical time-period at which the off-diagonal components in the $e-x$ sector grow, is  
$\tau_L \sim (\alpha\omega \mu^{\ast}_r)^{-1/2}$.
The mass hierarchy implies that $\tau_L\simeq 8 \tau_H$. As a consequence
of this slower growth of the $e-x$ conversions, the multi-angle
suppression is enough to make $\tau_L$ too slow to be effective.

 The suppression of conversions between  ${\nu}_e$ and ${\nu}_x$  explains the appearance  of  the high-energy spectral split in $\nu_e$ observed in the multi-angle case of Fig.~7. 
Concerning antineutrinos, we realize that not only $e-x$, but also $e-y$ conversions are strongly inhibited.
 We associate this behavior with the stronger violation of adiabaticity in the antineutrino sector, due to the smaller
spectral differences among  ${\bar\nu}_e$ and ${\bar\nu_x}$~\cite{Dasgupta:2010cd}.

We stress that the adiabaticity  plays a crucial role in determining
the impact of the multi-angle suppression  in the flavor evolution. 
Indeed, we explicitly checked that for the same flux ordering,  increasing the adiabaticity, by increasing the neutrino-neutrino potential by a factor of five,
the multi-angle suppression is less dramatic. In particular, $e-x$ conversions  reappear. 
Moreover, also the suppression of flavor oscillations in the antineutrino sector is 
relieved.
We find that this result is in agreement with the case shown in~\cite{Duan:2010bf}. 
There the choice of higher neutrino luminosities with respect to the ones
we are considering as benchmark case, allows the multi-angle suppression
to be strong only at low-radii, while the
final spectra are similar to the ones obtained in the single-angle case.

\begin{figure}[!t]
\begin{center}  
\includegraphics[width=\columnwidth]{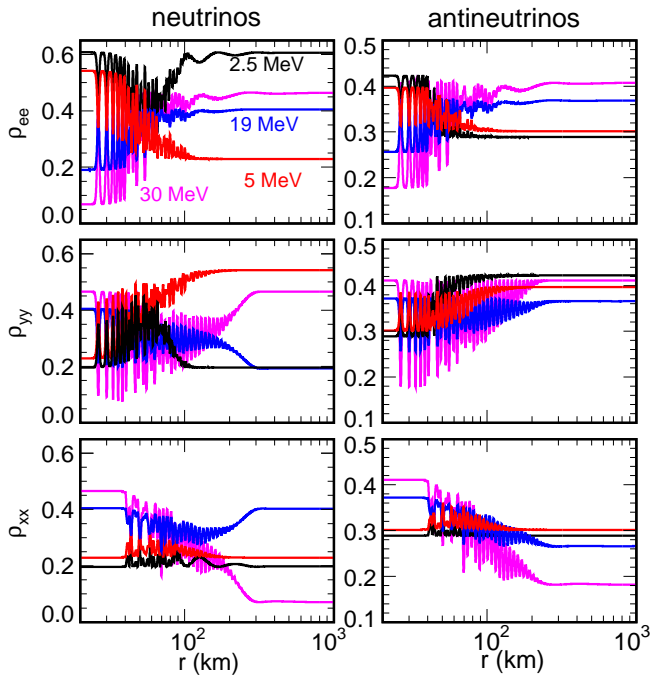}  
\end{center}  
\caption{Case   with $\Phi^0_{\nu_e}: \Phi^0_{{\bar{\nu}}_e}:
\Phi^0_{\nu_x} = 0.81:0.79:1.0$.
 Three-flavor evolution in inverted mass hierarchy in single-angle case.
Radial evolution of the diagonal components of the density matrix $\rho$ for neutrinos (left panels) and
antineutrinos (right panels)  for  different energy modes.
\label{fig:11}} 
\end{figure}  

\begin{figure}[!t]
\begin{center}  
\includegraphics[width=\columnwidth]{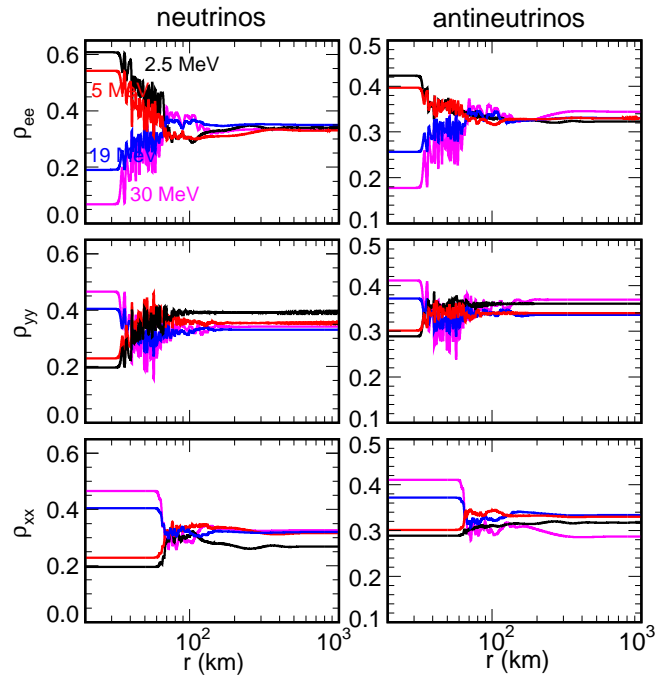}  
\end{center}  
\caption{As in Fig.~11, but for the
\emph{multi-angle} case.
\label{fig:12}} 
\end{figure}  

\subsection{Spectrum with small flavor asymmetries}

Finally, we consider the case $\Phi^0_{\nu_e}: \Phi^0_{{\bar{\nu}}_e}:
\Phi^0_{\nu_x} = 0.81:0.79:1.0$ which is intended to represent 
a case which a flux ordering possible at late times, where  asymmetries among 
$\nu_e$ and $\overline\nu_e$
can become small~\cite{Huedepohl:2009wh,Fischer:2009af}. The small asymmetry case has been pointed as representative
of flavor decoherence associated to multi-angle effects~\cite{Raffelt:2007yz,EstebanPretel:2007ec}. 

In Fig.~9  show the initial  (anti)neutrino  spectra for all the flavors and the final electron (anti)neutrino ones for the single-angle case (upper panels)
and the corresponding conversion probablities at the end of the flavor evolution.
Multi-angle results are shown in Fig.~10. The splitting features in the single-angle case are similar to the ones observed in Fig.~5 
of Sec. III~B.
However, the multi-angle results are dramatically different. We observe that neutrino and antineutrino spectra tend toward   flavor  equilibration with
different conversion probabilites displaced around $1/3$.
 
 In Fig.~11 and 12 we represent the radial evolution of the diagonal elements of the density matrix $\rho_{ee}$, $\rho_{yy}$, $\rho_{xx}$ for different energy modes for neutrinos (left panels) and antineutrinos (right panels). 
In the $\rho_{ee}$ panels the order of the energy modes is $E= 2.5, 5, 19, 30$~MeV
going from the  curve starting with the highest value to the lowest one. This order is reversed in the $\rho_{yy}$ and $\rho_{xx}$ panels.   
Figure~11 represents the single-angle case, while Fig.~12 represents the multi-angle case. The difference of the flavor evolution in the two cases is striking.
In particular, multi-angle suppression blocks $e$-$y$ flavor conversions till $r\simeq 40$~km, as predictable from 
Eq.~(\ref{eq:multisup}). Then,
since conversions start at small $r$ where the evolution is more adiabatic
than in the case of Sec. III~B, also 
  $e$-$x$ oscillations have chance to develop   at $r\gtrsim  60$~km.
The stronger adiabaticity in this case allow also
 multi-angle effects to  have enough time to develop before the neutrino-neutrino
interaction term becomes small. Then, the multi-angle  effects smear the flavor conversions,  producing 
a tendency towards a three-flavor decoherence of the ensemble in  both $\nu$ and ${\overline\nu}$ sectors.

\begin{table*}[!t]
\begin{center}
\caption{Summary of multi-angle effects, $3\nu$ effects and spectral splts for different SN neutrino fluxes.}
\begin{tabular}{cccc}
\hline \hline
Initial spectral pattern & Multi-angle effects   & $\Delta m^2_{\rm sol}$-effects & Spectral splits \vbox to12pt{}\\
\hline
Single crossing & marginal & 
 absent & robust 
\vbox to12pt{} \\
\hline
Multiple crossings & relevant & 
 present/absent & smeared
\vbox to12pt{} \\
\hline
Small flavor asymmetries & strong & 
 present & washed-out 
\vbox to12pt{} \\
\hline
\end{tabular}
\label{tab:nuebar-effects}
\end{center}
\end{table*}

 \section{Conclusions}

 We have performed an exploration on the dependence  of multi-angle effects in self-induced supernova neutrino 
oscillations on the original neutrino fluxes.  
  Most of the previous studies~\cite{Duan:2006an,Fogli:2007bk,Fogli:2008pt,EstebanPretel:2007ec}
focused on neutrino fluxes typical of the accretion phase, with a pronounced $\nu_e$ excess, 
\emph{de facto} behaving like spectra with a single crossing.
In this situation, the 
synchronization of different angular modes at low-radii  prevails over 
multi-angle effects. Then, when flavor conversions start, these  are  adiabatic 
to produce the spectral swaps and splits, but not enough to allow multi-angle decoherence
to emerge. The result is    the known ``quasi single-angle'' evolution. 
 
However, one has to be cautious in generalizing this reassuring result.
In this context, we have shown that multi-angle effects can produce significant
deviations in the flavor evolution with respect to the three-flavor single-angle case, for neutrino fluxes
with a moderate flavor hierarchy and a $\nu_x$ excess, as possbile  during the supernova cooling phase. In this situation, the presence of multiple crossing points in the original  neutrino spectra destabilizes the synchronization at large neutrino densities~\cite{Raffelt:2008hr}. In absence of  synchronization, 
in the single-angle scheme collective conversions would be possible at low-radii. However, 
 multi-angle effects introduce a large dispersion in the neutrino-neutrino potential,
that prevents any possible collective flavor conversion at low radii. As a consequence, in the multi-angle scheme there would be
a significant delay of  the onset of the flavor  conversions,
 as recently observed~\cite{Duan:2010bf}. 
 We have shown   that this multi-angle delay  can produce dramatic
 changes, not only in the deepest supernova regions, as 
shown in~\cite{Duan:2010bf}, but also  in the final oscillated  neutrino spectra.
In particular,  the
 multi-angle suppression can be so strong to allow the onset of the   flavor evolution only at a large radius, when
the evolution is less adiabatic. Depending on the violation of adiabaticity, in the inverted mass hierarchy
there could   be  a suppression
 of the three-flavor effects, associated to the  solar sector.
This would dramatically change the pattern of swaps and splits in the final 
neutrino spectra.  
Finally, if the flavor asymmetries between $\nu_e$ and ${\overline \nu}_e$ are very small, the 
multi-angle suppression occurs only close to the neutrinosphere.
 In this situation, the stronger adiabaticity of the evolution 
allows  multi-angle effects to act efficiently also after the onset of the conversions,  tending to establish a three-flavor equilibration 
in both neutrino and antineutrino sectors. 

In Table~I we summarize our results on the role of multi-angle effects, $3\nu$ effects and spectral splits 
for different SN neutrino fluxes.
In this work we have explicitly shown numerical results only for the case of neutrino inverted mass hierarchy.
However, we have checked that the impact of multi-angle effects is qualitatively similar also for the 
normal hierarchy case.

From our numerical explorations, it results that self-induced flavor transformations of supernova neutrinos during the cooling
phase are  a continuous  source of surprises. The richness of the  phenomenology,
in the presence of neutrino spectra with multiple crossing points, 
was first realized in~\cite{Dasgupta:2009mg} with the discovery of the possibility
of multiple spectral splits in both the mass hierarchies. Then, it was realized that
for these neutrino fluxes, three-flavor effects can play a significant role in inverted mass hierarchy, changing
the splitting pattern expected from two-flavor calculations~\cite{Dasgupta:2010cd,Friedland:2010sc}.
Now, we show that also multi-angle effects are crucial in characterizing  the flavor  evolution in this case,
and could potentially kill the three-flavor effects.
In general, self-induced flavor conversions for spectra with multiple crossing points challenge most of the naive expectations 
on which was based the original picture of the collective supernova  neutrino conversions: low-radii synchronization, subleading role of multi-angle
and three-flavor effects.

The discovery of these new effects adds additional layers of complications in the simulation of the flavor evolution 
for supernova neutrinos. In particular, our result shows that during the cooling phase three-flavor multi-angle simulations
are crucial to obtain a correct result. Multi-angle effects would  be taken into account to assess
the impact of collective neutrino oscillations on the r-process nucleosynthesis in supernovae,  as recently investigated
in~\cite{Duan:2010af}. 
The impact of the multi-angle effects would crucially depend on different SN input: neutrino luminosities and flavor asymmetries, neutrinosphere
radius, etc. Since all these quantities  significantly change during the neutrino emission, one would expect time-dependent 
effects. At this regard,
 the possibility to detect signatures of these effects  in the next galactic  
supernova neutrino  burst~\cite{Choubey:2010up} would motivate further analytical and numerical investigations.

\subsection*{Appendix}

We discuss here a few technical aspects of the multi-angle numerical simulations, we performed on our
local computer facility (with Fortran 77 codes running a Linux cluster with 48 processors per CPU with 128 Gb of
shared RAM memory). 
Equation~(\ref{eomMA}), after discretization, provides a set of $16\times N_E \times N_u$ ordinary differential equations
 in $r$, where $N_E$ and $N_u$ are the number of points sampling
the (anti)neutrino energy $E$ and emission angle. In particular,
we find convenient to label the neutrino angular modes in terms of the variable~\cite{EstebanPretel:2007ec}
\begin{equation}
u = \sin^2 \theta_R \,\ ,
\end{equation}
  where $\theta_R$ is the zenith angle at the neutrino sphere $r = R$
of a given mode relative to the radial direction. With this choice, the parameter
$u$ is fixed for every neutrino trajectory.

We have then performed our  simulations of the three-flavor neutrino  evolution using  a Runge-Kutta integration
routine taken from the CERNLIB libraries~\cite{cernlib}. 
We fixed the numerical tolerance of the integrator at the level of $10^{-6}$ and 
 increased the number of sampling points in angle and energy till we reach a stable numerical result.
In this situation  
we estimate a numerical (fractional)
accuracy of our results better than $10^{-2}$. 
In  order to have a clear energy resolution of the spectral splits we 
took $N_E= 10^2$ energy points, equally distributed in the range $E \in [0.1,80]$~MeV.
The number of angular modes is also a crucial choice, since it is well known that a sparse sampling
in angle can lead to numerical artifacts that   would destroy the collective behavior of the neutrino
self-induced conversions~\cite{EstebanPretel:2007ec}. 
Typically, numerical stability would require
$N_u = {\mathcal O}(10^3)$ angular modes.
  
 Since we can claim to have reached stable numerical simulations,  
we are confident in the accuracy of
the results obtained in this work. Moreover, we have been
  able to reproduce previous results presented in literature for cases
similar to the ones we are investigating (see, e.g.,~\cite{Fogli:2007bk,Duan:2010bf}).

\begin{acknowledgments}  
We thank Sovan Chakraborty, Basudeb Dasgupta and Georg Raffelt for reading
the manuscript and for comments on it.
  The work of A.M. was supported by the German Science Foundation (DFG)
within the Collaborative Research Center 676 ``Particles, Strings and the
Early Universe''.
\end{acknowledgments}



\begin{thebibliography}{00}   


  
 \bibitem{Matt}  L.~Wolfenstein,  
				``Neutrino Oscillations In Matter,''  
                Phys.\ Rev.\ D {\bf 17}, 2369 (1978);  
                S. P.~Mikheev and A. Yu.\ Smirnov,  
                ``Resonance Enhancement Of Oscillations In Matter And Solar Neutrino  
				Spectroscopy,''  
                Yad.\ Fiz.\ {\bf 42}, 1441 (1985)  
                [Sov.\ J.\ Nucl.\ Phys.\ {\bf 42}, 913 (1985)].  



\bibitem{Dighe:1999bi}
  A.~S.~Dighe and A.~Y.~Smirnov,
  ``Identifying the neutrino mass spectrum from the neutrino burst from a
  supernova,''
  Phys.\ Rev.\  D {\bf 62}, 033007 (2000)
  [hep-ph/9907423].


 
\bibitem{Pantaleone:1992eq}  
  J.~Pantaleone,  
  ``Neutrino oscillations at high densities,''  
  Phys.\ Lett.\ B {\bf 287}, 128 (1992).  
  
\bibitem{Sigl:1992fn}  
  G.~Sigl and G.~Raffelt,  
  ``General kinetic description of relativistic mixed neutrinos,''  
  Nucl.\ Phys.\ B {\bf 406}, 423 (1993).  
  
    
\bibitem{McKellar:1992ja}
  B.~H.~J.~McKellar and M.~J.~Thomson,
  ``Oscillating doublet neutrinos in the early universe,''
  Phys.\ Rev.\  D {\bf 49}, 2710 (1994).
  
    
\bibitem{Qian:1994wh}
  Y.~Z.~Qian and G.~M.~Fuller,
  ``Neutrino-neutrino scattering and matter enhanced neutrino flavor
  transformation in Supernovae,''
  Phys.\ Rev.\  D {\bf 51}, 1479 (1995)
  [:astro-ph/9406073].
  
\bibitem{Samuel:1993uw}  
  S.~Samuel,  
  ``Neutrino oscillations in dense neutrino gas\-es,''  
  Phys.\ Rev.\ D {\bf 48}, 1462 (1993).  
  
\bibitem{Kostelecky:1993dm}  
  V.~A.~Kosteleck\'y and S.~Samuel,  
  ``Neutrino oscillations in the early universe with an inverted  
  neutrino mass hierarchy,''  
  Phys.\ Lett.\ B {\bf 318}, 127 (1993).  
  
\bibitem{Kostelecky:1995dt}  
  V.~A.~Kosteleck\'y and S.~Samuel,  
  ``Self-maintained coherent oscillations in dense neutrino gases,''  
  Phys.\ Rev.\ D {\bf 52}, 621 (1995)  
  [hep-ph/9506262].  
  
\bibitem{Samuel:1996ri}  
  S.~Samuel,  
  ``Bimodal coherence in dense selfinteracting neutrino gases,''  
  Phys.\ Rev.\ D {\bf 53}, 5382 (1996)  
  [hep-ph/9604341].  



  
\bibitem{Pastor:2001iu}  
  S.~Pastor, G.~G.~Raffelt and D.~V.~Semikoz,  
  ``Physics of synchronized neutrino oscillations caused by  
  self-interactions,''  
  Phys.\ Rev.\ D {\bf 65}, 053011 (2002)  
  [hep-ph/0109035].  
  
\bibitem{Wong:2002fa}  
  Y.~Y.~Y.~Wong,  
  ``Analytical treatment of neutrino asymmetry equilibration  
  from flavour oscillations in the early universe,''  
  Phys.\ Rev.\ D {\bf 66}, 025015 (2002)  
  [hep-ph/ 0203180].  
  
\bibitem{Abazajian:2002qx}  
  K.~N.~Abazajian, J.~F.~Beacom and N.~F.~Bell,  
  ``Stringent constraints on cosmological neutrino antineutrino asymmetries  
  from synchronized flavor transformation,''  
  Phys.\ Rev.\  D {\bf 66}, 013008 (2002)  
  [arXiv:astro-ph/0203442].  
  
\bibitem{Pastor:2002we}  
  S.~Pastor and G.~Raffelt,  
  ``Flavor oscillations in the supernova hot bubble region:  
  Nonlinear  effects of neutrino background,''  
  Phys.\ Rev.\ Lett.\  {\bf 89}, 191101 (2002)  
  [astro-ph/0207281].  
  
\bibitem{Sawyer:2004ai}  
  R.~F.~Sawyer,  
  ``Classical instabilities and quantum speed-up in the evolution of  
  neutrino clouds,''  
  hep-ph/0408265.  
  
\bibitem{Sawyer:2005jk}  
  R.~F.~Sawyer,  
  ``Speed-up of neutrino transformations in a supernova environment,''  
  Phys.\ Rev.\  D {\bf 72}, 045003 (2005)  
  [hep-ph/0503013].  


\bibitem{Duan:2005cp}
  H.~Duan, G.~M.~Fuller and Y.~Z.~Qian,
  ``Collective Neutrino Flavor Transformation In Supernovae,''
  Phys.\ Rev.\  D {\bf 74}, 123004 (2006)
  [astro-ph/0511275].


\bibitem{Duan:2006an}
  H.~Duan, G.~M.~Fuller, J.~Carlson and Y.~Z.~Qian,
  ``Simulation of coherent non-linear neutrino flavor transformation in the
  supernova environment. I: Correlated neutrino trajectories,''
  Phys.\ Rev.\  D {\bf 74}, 105014 (2006)
  [astro-ph/0606616].
  
\bibitem{Hannestad:2006nj}
  S.~Hannestad, G.~G.~Raffelt, G.~Sigl and Y.~Y.~Y.~Wong,
  ``Self-induced conversion in dense neutrino gases: Pendulum in flavour  
 space,''
  Phys.\ Rev.\  D {\bf 74}, 105010  (2006)
  [Erratum-ibid.\  D {\bf 76},  029901 (2007)]
  [astro-ph/0608695].


\bibitem{Duan:2010bg}
  H.~Duan, G.~M.~Fuller and Y.~Z.~Qian,
  ``Collective Neutrino Oscillations,''
  arXiv:1001.2799 [hep-ph].





\bibitem{Fogli:2007bk}
  G.~L.~Fogli, E.~Lisi, A.~Marrone and A.~Mirizzi,
  ``Collective neutrino flavor transitions in supernovae and the role of
  trajectory averaging,''
  JCAP {\bf 0712}, 010 (2007)
  [arXiv:0707.1998 [hep-ph]].

\bibitem{Fogli:2008pt}
  G.~L.~Fogli, E.~Lisi, A.~Marrone, A.~Mirizzi and I.~Tamborra,
  ``Low-energy spectral features of supernova (anti)neutrinos in inverted
  hierarchy,''
  Phys.\ Rev.\  D {\bf 78}, 097301 (2008)
  [arXiv:0808.0807 [hep-ph]].



\bibitem{Raffelt:2007cb}
  G.~G.~Raffelt and A.~Y.~Smirnov,
  ``Self-induced spectral splits in supernova neutrino fluxes,''
  Phys.\ Rev.\  D {\bf 76}, 081301 (2007)
  [Erratum-ibid.\  D {\bf 77}, 029903 (2008)]
  [arXiv:0705.1830 [hep-ph]].


\bibitem{Raffelt:2007xt}
  G.~G.~Raffelt, A.~Y.~Smirnov,
  ``Adiabaticity and spectral splits in collective neutrino transformations,''
  Phys.\ Rev.\  {\bf D76}, 125008 (2007).
  [arXiv:0709.4641 [hep-ph]].

\bibitem{Duan:2007bt}
  H.~Duan, G.~M.~Fuller, J.~Carlson {\it et al.},
  ``Neutrino Mass Hierarchy and Stepwise Spectral Swapping of Supernova Neutrino Flavors,''
  Phys.\ Rev.\ Lett.\  {\bf 99}, 241802 (2007).
  [arXiv:0707.0290 [astro-ph]].


\bibitem{Duan:2008za}
  H.~Duan, G.~M.~Fuller, Y.~-Z.~Qian,
  ``Stepwise spectral swapping with three neutrino flavors,''
  Phys.\ Rev.\  {\bf D77}, 085016 (2008).
  [arXiv:0801.1363 [hep-ph]].


\bibitem{Gava:2008rp}
  J.~Gava and C.~Volpe,
  ``Collective neutrinos oscillation in matter and CP-violation,''
  Phys.\ Rev.\  D {\bf 78}, 083007 (2008)
  [arXiv:0807.3418 [astro-ph]].


\bibitem{Gava:2009pj}
  J.~Gava, J.~Kneller, C.~Volpe and G.~C.~McLaughlin,
  ``A dynamical collective calculation of supernova neutrino signals,''
  Phys.\ Rev.\ Lett.\  {\bf 103}, 071101 (2009)
  [arXiv:0902.0317 [hep-ph]].


\bibitem{Dasgupta:2009mg}
  B.~Dasgupta, A.~Dighe, G.~G.~Raffelt and A.~Y.~Smirnov,
  ``Multiple Spectral Splits of Supernova Neutrinos,''
  Phys.\ Rev.\ Lett.\  {\bf 103}, 051105 (2009)
  [arXiv:0904.3542 [hep-ph]].




\bibitem{Friedland:2010sc}
  A.~Friedland,
  ``Self-refraction of supernova neutrinos: mixed spectra and three-flavor
  instabilities,''
  Phys.\ Rev.\ Lett.\  {\bf 104}, 191102 (2010)
  [arXiv:1001.0996 [hep-ph]].

\bibitem{Dasgupta:2010cd}
  B.~Dasgupta, A.~Mirizzi, I.~Tamborra and R.~ Tom{\`a}s,
  ``Neutrino mass hierarchy and three-flavor spectral splits of supernova
  neutrinos,''
  Phys.\ Rev.\  D {\bf 81}, 093008 (2010)
  [arXiv:1002.2943 [hep-ph]].

\bibitem{Fogli:2009rd}
  G.~Fogli, E.~Lisi, A.~Marrone and I.~Tamborra,
  ``Supernova neutrinos and antineutrinos: ternary luminosity diagram and
  spectral split patterns,''
  JCAP {\bf 0910}, 002 (2009)
  [arXiv:0907.5115 [hep-ph]].

\bibitem{Choubey:2010up}
  S.~Choubey, B.~Dasgupta, A.~Dighe and A.~Mirizzi,
  ``Signatures of collective and matter effects on supernova neutrinos at large
  detectors,''
  arXiv:1008.0308 [hep-ph].

\bibitem{Pantaleone:1992xh}
  J.~T.~Pantaleone,
  ``Dirac neutrinos in dense matter,''
  Phys.\ Rev.\  D {\bf 46}, 510 (1992).



\bibitem{Raffelt:2007yz}
  G.~G.~Raffelt and G.~Sigl,
  ``Self-induced decoherence in dense neutrino gases,''
  Phys.\ Rev.\  D {\bf 75}, 083002 (2007)
  [hep-ph/0701182].
  
  
  
\bibitem{EstebanPretel:2007ec}
  A.~Esteban-Pretel, S.~Pastor, R.~ Tom{\`a}s, G.~G.~Raffelt and G.~Sigl,
  ``Decoherence in supernova neutrino transformations suppressed by
  deleptonization,''
  Phys.\ Rev.\  D {\bf 76}, 125018 (2007)
  [arXiv:0706.2498 [astro-ph]].

\bibitem{Sawyer:2008zs}
  R.~F.~Sawyer,
  ``The multi-angle instability in dense neutrino systems,''
  Phys.\ Rev.\  D {\bf 79} (2009) 105003
  [arXiv:0803.4319 [astro-ph]].
  
\bibitem{Raffelt:2003en}
  G.~G.~Raffelt, M.~T.~Keil, R.~Buras, H.~T.~Janka and M.~Rampp,
  ``Supernova neutrinos: Flavor-dependent fluxes and spectra,''
 astro-ph/0303226.

\bibitem{Fischer:2009af}
  T.~Fischer, S.~C.~Whitehouse, A.~Mezzacappa {\it et al.},
  ``Protoneutron star evolution and the neutrino driven wind in general relativistic neutrino radiation hydrodynamics simulations,''
  Astron.\ Astrophys.\  {\bf 517}, A80 (2010).
  [arXiv:0908.1871 [astro-ph.HE]].

\bibitem{Huedepohl:2009wh}
  L.~Hudepohl, B.~Muller, H.~-T.~Janka {\it et al.},
  ``Neutrino Signal of Electron-Capture Supernovae from Core Collapse to Cooling,''
  Phys.\ Rev.\ Lett.\  {\bf 104}, 251101 (2010).
  [arXiv:0912.0260 [astro-ph.SR]].
 
\bibitem{Cherry:2010yc}
  J.~F.~Cherry, G.~M.~Fuller, J.~Carlson, H.~Duan and Y.~Z.~Qian,
  ``Multi-Angle Simulation of Flavor Evolution in the Neutrino Neutronization
  Burst From an O-Ne-Mg Core-Collapse Supernova,''
  Phys.\ Rev.\  D {\bf 82} (2010) 085025
  [arXiv:1006.2175 [astro-ph.HE]].

  
  
\bibitem{Duan:2010bf}
  H.~Duan and A.~Friedland,
  ``Self-induced suppression of collective neutrino oscillations in a
  supernova,''   Phys.\ Rev.\ Lett.\  {\bf 106}, 091101 (2011)
  [arXiv:1006.2359 [hep-ph]].

\bibitem{Raffelt:2008hr}
  G.~G.~Raffelt,
  ``Self-induced parametric resonance in collective neutrino oscillations,''
  Phys.\ Rev.\  D {\bf 78} (2008) 125015
  [arXiv:0810.1407 [hep-ph]].

\bibitem{animations}
Animated figures at 
http ://www.mppmu.mpg.de/supernova/ multisplits/
  

\bibitem{Dasgupta:2007ws}
  B.~Dasgupta and A.~Dighe,
  ``Collective three-flavor oscillations of supernova neutrinos,''
  Phys.\ Rev.\  D {\bf 77}, 113002 (2008)
  [arXiv:0712.3798 [hep-ph]].

\bibitem{Keil:2002in}
  M.~T.~Keil, G.~G.~Raffelt and H.~T.~Janka,
  ``Monte Carlo study of supernova neutrino spectra formation,''
  Astrophys.\ J.\  {\bf 590}, 971 (2003)
  [astro-ph/0208035].

  





\bibitem{Dasgupta:2008cu}
  B.~Dasgupta, A.~Dighe, A.~Mirizzi and G.~G.~Raffelt,
  ``Collective neutrino oscillations in non-spherical geometry,''
  Phys.\ Rev.\  D {\bf 78}, 033014 (2008)
  [arXiv:0805.3300 [hep-ph]].

\bibitem{Duan:2008fd}
  H.~Duan, G.~M.~Fuller, Y.~-Z.~Qian,
  ``Symmetries in collective neutrino oscillations,''
  J.\ Phys.\ G {\bf G36}, 105003 (2009).
  [arXiv:0808.2046 [astro-ph]].



 
\bibitem{GonzalezGarcia:2010er}
  M.~C.~Gonzalez-Garcia, M.~Maltoni and J.~Salvado,
  ``Updated global fit to three neutrino mixing: status of the hints of $\theta_{13} > 0$,''
  JHEP {\bf 1004} (2010) 056
  [arXiv:1001.4524 [hep-ph]].
  
  

  

 
  
\bibitem{EstebanPretel:2008ni}
  A.~Esteban-Pretel, A.~Mirizzi, S.~Pastor, R.~ Tom{\`a}s, G.~G.~Raffelt, P.~D.~Serpico and G.~Sigl,
  ``Role of dense matter in collective supernova neutrino transformations,''
  Phys.\ Rev.\  D {\bf 78}, 085012 (2008)
  [arXiv:0807.0659 [astro-ph]].

  
  
\bibitem{Dasgupta:2008cd}
  B.~Dasgupta, A.~Dighe, A.~Mirizzi and G.~G.~Raffelt,
  ``Spectral split in prompt supernova neutrino burst: Analytic three-flavor treatment,''
  Phys.\ Rev.\  {\bf D77}, 113007 (2008).
  [arXiv:0801.1660 [hep-ph]].


\bibitem{Fogli:2008fj}
  G.~Fogli, E.~Lisi, A.~Marrone and I.~Tamborra,
  ``Supernova neutrino three-flavor evolution with dominant collective effects,''
  JCAP {\bf 0904}, 030 (2009).
  [arXiv:0812.3031 [hep-ph]].


\bibitem{Duan:2010af}
  H.~Duan, A.~Friedland, G.~C.~McLaughlin and R.~Surman,
  ``The influence of collective neutrino oscillations on a supernova
  r-process,''
  arXiv:1012.0532 [astro-ph.SR].


\bibitem{cernlib} 
http://dollywood.itp.tuwien.ac.at/cernlib/



\end{thebibliography}
\end{document}